\documentclass[sigconf,natbib=true,anonymous=false,screen=true]{acmart}

\usepackage{color}
\usepackage{flushend}
\usepackage{graphicx}
\usepackage{booktabs}
\usepackage[htt]{hyphenat}
\usepackage[]{xcolor}
\usepackage{tabularx}
\usepackage{booktabs}
\usepackage{multirow}
\usepackage[inline]{enumitem}
\usepackage{xspace}
\usepackage{acronym}

\definecolor{midnightgreen}{rgb}{0.0, 0.29, 0.33}

\newcommand{\RG}{R$\rightarrow$G\xspace}
\newcommand{\GRG}{G$\rightarrow$R$\rightarrow$G\xspace}

\acrodef{CSA}{Conversational Search Agent}
\acrodef{PTKB}{Personal Text Knowledge Base}
\acrodef{TREC}{TExt Retrieval Conference}
\acrodef{iKAT}{Interactive Knowledge Assistance Track}
\acrodef{CAsT}{Conversational Assistance Track}
\acrodef{NIST}{National Institute of Standards and Technology}
\acrodef{LLM}{Large Language Model}
\copyrightyear{2024} 
\acmYear{2024} 
\setcopyright{acmlicensed}\acmConference[SIGIR '24]{Proceedings of the 47th International ACM SIGIR Conference on Research and Development in Information Retrieval}{July 14--18, 2024}{Washington, DC, USA}
\acmBooktitle{Proceedings of the 47th International ACM SIGIR Conference on Research and Development in Information Retrieval (SIGIR '24), July 14--18, 2024, Washington, DC, USA}
\acmDOI{10.1145/3626772.3657860}
\acmISBN{979-8-4007-0431-4/24/07}

\begin{document}

\title[TREC iKAT 2023]{TREC iKAT 2023: A Test Collection for Evaluating Conversational and Interactive Knowledge Assistants}

\author{Mohammad Aliannejadi}
\orcid{0000-0002-9447-4172} 
\affiliation{%
  \institution{University of Amsterdam}
  \city{Amsterdam}
  \country{The Netherlands}
}
\email{m.aliannejadi@uva.nl}

\author{Zahra Abbasiantaeb}
\orcid{0000-0002-4046-3419} 
\affiliation{%
  \institution{University of Amsterdam}
  \city{Amsterdam}
  \country{The Netherlands}
}
\email{z.abbasiantaeb@uva.nl}

\author{Shubham Chatterjee}
\orcid{0000-0002-6729-1346}
\affiliation{%
  \institution{University of Edinburgh}
  \city{Edinburgh}
  \country{Scotland, UK}
}
\email{shubham.chatterjee@ed.ac.uk}

\author{Jeffery Dalton}
\orcid{0000-0003-2422-8651}
\affiliation{%
  \institution{University of Edinburgh}
  \city{Edinburgh}
  \country{Scotland, UK}
}
\email{jeff.dalton@ed.ac.uk}

\author{Leif Azzopardi}
\orcid{0000-0002-6900-0557}
\affiliation{%
  \institution{University of Strathclyde}
  \city{Glasgow}
  \country{Scotland, UK}
}
\email{leif.azzopardi@strath.ac.uk}

\renewcommand{\shortauthors}{Aliannejadi et al.}
\begin{abstract}

Conversational information seeking has evolved rapidly in the last few years with the development of \acp{LLM}, providing the basis for interpreting and responding in a naturalistic manner to user requests.
The extended TREC \ac{iKAT} collection aims to enable researchers to test and evaluate their \ac{CSA}. The collection contains a set of  36 personalized dialogues over 20 different topics each coupled with a \ac{PTKB} that defines the bespoke user personas. A total of 344 turns with approximately 26,000 passages are provided as assessments on relevance, as well as additional assessments on generated responses over four key dimensions: relevance, completeness, groundedness, and naturalness. 
The collection challenges \acp{CSA} to efficiently navigate diverse personal contexts, elicit pertinent persona information, and employ context for relevant conversations.

The integration of a \ac{PTKB} and the emphasis on decisional search tasks contribute to the uniqueness of this test collection, making it an essential benchmark for advancing research in conversational and interactive knowledge assistants.  
\end{abstract}
\begin{CCSXML}
<ccs2012>
   <concept>
       <concept_id>10003120.10003123.10011758</concept_id>
       <concept_desc>Human-centered computing~Interaction design theory, concepts and paradigms</concept_desc>
       <concept_significance>500</concept_significance>
       </concept>
 </ccs2012>
\end{CCSXML}
\ccsdesc[500]{Human-centered computing~Interaction design theory, concepts and paradigms}

\keywords{Conversational Information Seeking, Conversational Search Agents, Evaluation, Test Collection}

\maketitle              
\begin{figure}[!b]
    \centering
    \includegraphics[width=\columnwidth]{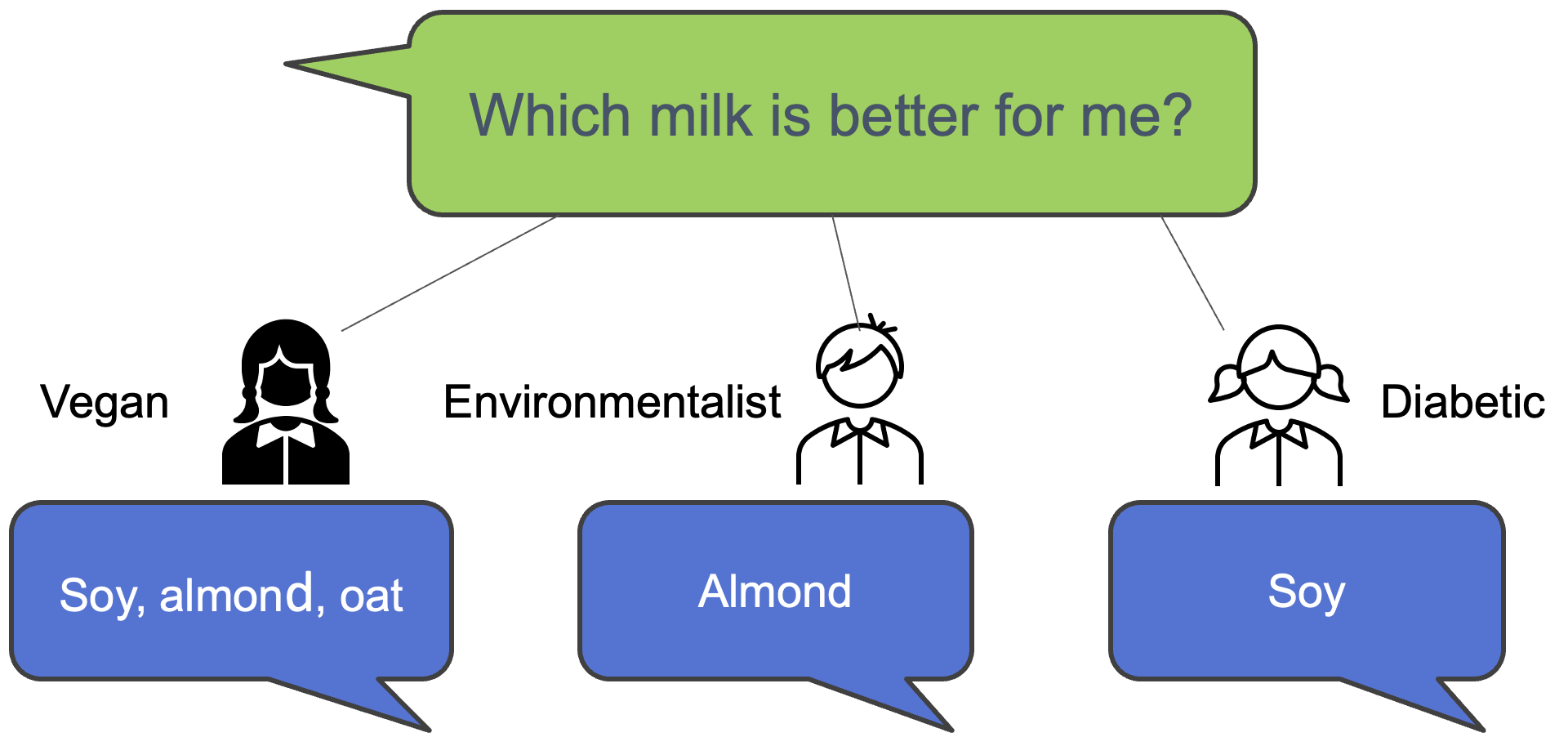}
    \caption{Example outcomes given a conversation on alternatives to cow's milk with three different personas.}
    \label{fig:milk}
\end{figure}

\section{Introduction}

Conversational information seeking provides a natural and intuitive way for users to interact and discover relevant information through dialogue with an agent~\cite{radlinski2017theoretical, azzopardi2018conceptualizing}. With the advent of \acfp{LLM}~\cite{devlin2019bert}, access to \acfp{CSA} has become a reality (e.g., BingChat, Bard, BlenderBot, etc.). Moreover, the underlying technology has become sufficiently accessible, to afford the wide-scale research and development of such agents. However, resources to evaluate \acp{CSA} are currently limited. While numerous test collections and resources exist to test \acp{LLM} over a variety of different tasks, the conversational information-seeking task presents numerous varied and complex evaluation challenges~\cite{DBLP:conf/eacl/RahmaniWANY24,DBLP:conf/wsdm/Abbasiantaeb0KA24} for a number of reasons.

\begin{itemize}%
    \item Conversations are context-dependent (i.e., past turns matter).
    \item Conversations are personal (i.e., user preferences matter).
    \item Conversations can evolve in many ways, shapes, and forms (i.e., conversational flow matters).
    \item Conversations can involve mixed initiative (i.e., user-agent interplay matters). 
\end{itemize}

This means that when a \ac{CSA} is responding to a request (in the context of the conversation), the same request (question) might yield considerably different responses (and answers), based on what has been previously uttered, and be contingent on the user's preferences. For example, consider the following scenario where the topic is about \textit{alternatives to cow's milk}. There are many different motivations why a person may be interested in milk alternatives, and which alternatives are explored and discussed depend on this motivation and the conversational history. Three personas (shown in Figure~\ref{fig:milk}) can illustrate this: 
\begin{itemize}%
    \item (A) Alice is a strict vegan who wants to find an alternative that is healthy but doesn't come from animals.
    \item (B) Bob is an environmentalist and wants to find an alternative that is high in calcium and minimizes harm to the environment.
    \item (C) Charlie has been recently diagnosed with diabetes and wants to find an alternative that is low in sugar.
\end{itemize}
Given Alice, Bob, and Charlie's ``personas,'' their conversations with an agent would evolve and develop in very different ways.
This is because what is relevant to one, may not be relevant to another. Relevance is conditional and contextually dependent. Consequently, by the end of their conversation, what they have learned about, what they have understood, and what they have decided regarding milk alternatives would vary, reflecting their personalized contexts.

The challenge lies in enabling \ac{CSA} to incorporate this personalized context to guide users effectively, considering the relevant information about the user. To create a resource that enables the evaluation of \ac{CSA}, TREC \acf{iKAT} was introduced in 2023. TREC \ac{iKAT} also emphasizes decisional search tasks~\cite{russell2018information}, where users need to sift through data and information to weigh up options to reach a conclusion or perform an action (such as which milk to consume). These tasks are prevalent in everyday information seeking -- be it related to travel, health, or shopping --  they often revolve around a subset of high-level information operators where queries or questions about the information space include: finding options, comparing options, identifying the pros and cons of options, etc.~\cite{azzopardi2018conceptualizing}. Given the different personas and their information needs (expressed through the sequence of questions), diverse conversation trajectories will arise --- because the answers to these similar queries will be very different. 

In this paper, we describe the TREC iKAT track and present our extensions to the test collection, which provides a re-usable resource to the community for evaluating CSAs -- that enables researchers to consider research questions such as:

\begin{itemize}%
    \item \textbf{Dependent Relevance}: can agents effectively employ context and prior responses to foster relevant conversations?
    \item \textbf{Elicitation}: are agents proficient in drawing out pertinent persona information to customize discussions?
    \item \textbf{Personalization}: can agents provide tailored and relevant conversational responses based on the user's persona and history?
\end{itemize}
We present the performance results of TREC submissions along with additional baseline comparisons, analyzing the outcomes from multiple perspectives. This approach helps illustrate the usability of the proposed resources and highlights their limitations.

\section{Related Work}
The evaluation of Interactive Information Retrieval (IIR) systems and agents remains a persistent challenge within the broader field of information retrieval. The complexities arise from the dynamic, context-dependent, and personalized nature of user-system interactions, requiring ongoing research to develop effective evaluation methodologies~\cite{DBLP:conf/trec/BelkinCCKNPPRSX96,DBLP:conf/trec/Hersh02,belkin2008iir}. Over the years, various test collections have been developed to simulate parts of the search process (e.g., TREC Interactive Track 1998-2002~\cite{over2001trec} and TREC Dynamic Domain Track 2015-2017~\cite{DBLP:conf/trec/0001FS15,yang2016trec,DBLP:conf/trec/YangTS17}). However, these resources focus on document retrieval over rounds of feedback, rather than conversation. Recently, the TREC \ac{CAsT}~\cite{dalton2019cast,dalton2020cast,dalton2021cast,owoicho2023trec} has provided resources for conversational search tasks.

The first year of the \ac{CAsT} began with predetermined conversational trajectories and responses. These became longer and richer in the second year with the addition of dependence on system responses.  The third year increased result dependencies and added richer types of interactions including user feedback, as well as elementary forms of user revealment. However, the ability to have realistic interaction was limited. The fourth year of \ac{CAsT} aimed to add more conversational elements to the interaction streams, by introducing mixed initiatives (clarifications, and suggestions) to create multi-path and multi-turn conversations for each topic.

TREC iKAT evolved \ac{CAsT} into a new track to signal this new trajectory~\cite{aliannejadi2024trec}. TREC iKAT aims to focus on supporting multi-path, multi-turn, and multi-perspective conversations. That is for a given topic, the direction and the conversation that evolves depends not only on the prior responses but also on the persona of the user. We describe this track in more detail in the next section. \citet{DBLP:journals/tois/VakulenkoKR21} in their large-scale analysis of conversational datasets highlight the limitations of existing data collections, emphasizing the need to have personalized and knowledge-intensive dialogues where system--user interactions can be taken into account. TREC iKAT aims to address some of these limitations, where the introduction of a personal knowledge base and the inclusion of complex and knowledge-intensive dialogues pose several novel challenges.

A vast array of different resources exist to evaluate \acp{LLM} over a number of different tasks for a range of purposes, such as BigBench aims to test \acp{LLM} Intelligence and Capabilities \cite{srivastava2023imitation}; LaMDA aims to test LLMs linguistic precision on grammar~\cite{thoppilan2022lamda};
GLUE/SuperGLUE assess comprehension and dialogue tasks emphasizing understanding of intricate sentences for natural language processing~\cite{wang2020superglue};
HellaSwag aims to evaluate the common sense reasoning of LLMs~\cite{zellers2019hellaswag};
OpenAI provides Moderation, Reliability, and Fairness API to test the safety of models to ensure they filter out potentially harmful content, along with testing model bias and diversity; ParlAI\footnote{\url{https://parl.ai/}} provides a system for training and evaluating dialogues and chats;
Stanford's CoQA tests the comprehension of texts, and answering linked questions conversationally~\cite{reddy2019coqa}; Holistic Evaluation of Language Models~\cite{liang2023holistic} (HeLM) which provides a number of open domain and specialized domain one-shot question and answering tasks (where responses are short answers or multiple choice); and SQuAD tests reading comprehension using text segment answers from articles~\cite{rajpurkar2016squad}. In contrast to these test collections, and the many more that have been developed (see \citet{guo2023evaluating} for a recent survey on evaluating LLMs), the TREC iKAT resource aims to provide tasks within the context of personas -- for which the agent needs to help the user make a decision (or lead to making a decision).
\section{The iKAT Resources}

The iKAT resources presented here extend the TREC iKAT 2023 test collection \citet{aliannejadi2024trec}). 
In this work, we extend the TREC iKAT 2023 resources by providing human and automatic annotations of the generated responses, as well as further analysis of the submissions based on these annotations.
To contextualize the resources, we first provide an overview of the track.

The primary focus of the TREC iKAT is to challenge conversational search agents to deliver a relevant and informative response given the user's current request, past conversations, and their \acp{PTKB}, grounded on the results from the test collection.
While these responses by the agents can be extracted passages from the collection, agents can also amalgamate or summarize passages when generating a response. A requirement imposed on responses, though, is that they should cite at least one ``provenance'' passage from the collection, maintaining a focus on passage/provenance ranking (similar to TREC CAsT). The agents should also consider the previous conversational turns as context, equivalent to taking the parents in the conversational topic tree. Moreover, the agents should utilize the personal information provided in the \ac{PTKB}, which is a set of narrative sentences, when forming responses. The sentences are assumed to be collected from previous conversations of the user with the system --- which capture and describe the user's preferences, for example, that they prefer pizza to pasta, sunny weather to cold weather, etc. The collection contains twenty topics and a topic is associated with one to three distinct personas, providing 36 personalized conversational dialogues (see below for more details).

Given the focus, the main evaluation tasks are as follows where, for a given conversational topic, the context/history of the conversation, and the user's current utterance:

\begin{itemize}%
    \item \textbf{PTKB Statement Ranking Task:} the system should return a ranking of the PTKB statements based on their relevance to the current conversational turn. 
    \item \textbf{Passage Ranking Task:} the system should return a ranking of the passages from the collection based on their relevance to the conversation. 
    \item \textbf{Response Generation Task:} the system should return a response that provides the answer which is intended to be shown to the user. It should be fluent, satisfy their information need, and not contain extraneous or redundant information. The response could be a generative or abstractive summary of the relevant passages.
\end{itemize}

\subsection{Personalized Conversational Topics}
\label{sec:topics}

The TREC iKAT 2023 has 36 personalized conversational dialogues (11 for training and 25 for testing) over 20 different topics. For each topic, there are up to three different personas associated with them. For example, Topic 9 is on ``Finding a Diet,'' where we have two personas: (9-1) which characterizes a vegetarian who has medical conditions and allergies, and (9-2) which characterizes a middle-aged man who is overweight and has a knee injury. Each dialogue contains a number of user-system turns, where the average length of a dialogue is 13 turns (and 427 turns in total over all dialogues both in the training and test sets.). Not every turn necessitates referencing the statements in the PTKB. Therefore, only a subset of turns are classified as personalized turns—specifically those that include at least one relevant statement from the PTKB.
For these turns, the system must consider the PTKB information in order to answer the user's utterance accurately.

Guidelines for creating topics were established as follows: Initially, a PTKB (Personalized Topic Knowledge Base) was constructed for the designated topic. Using the associated persona (along with any previous responses or history), the next user utterance was generated. Subsequently, relevant statements from the PTKB were identified, leading to a search for pertinent passages. Once a sufficient number of relevant passages were located, a canonical response was formulated. This entire process was managed by the organizers (for detailed information on topic creation, refer to ~\cite{aliannejadi2024trec}).

In the development of the PTKBs, particular attention was paid to including only high-level personal information while excluding any personally identifiable information to safeguard the privacy of the contributors.

\subsection{Collection}
\label{sec:collection}
The collection used for passage retrieval is a subset of ClueWeb22-B~\cite{overwijk2022clueweb22}. 
To create this subset, we manually inspected the domains of the documents within the ClueWeb22-B dataset. We prioritized the diversity of domains and eliminated those that were not relevant. 
The final subset contains 116,838,987 passages, which was distributed by the Lemur Project.\footnote{\url{https://lemurproject.org/clueweb22/}} \looseness=-1

To segment the documents into passages, we used a similar methodology as the one used by the TREC Deep Learning track for MS MARCO. We performed the following steps:

\begin{enumerate}%
    \item Each document was initially shortened to a length of 10,000 characters.
    \item A sliding window approach was then used, where we took 10 consecutive sentences as a single passage.
    \item After these 10 sentences, we moved the window by 5 sentences (i.e., a 5-sentence stride) to create the next passage.
\end{enumerate}

\subsection{Assessments for PTKB Statement Relevance Task}
\label{sec:ptkbRelevance}
To assess the relevance of PTKB statements for each turn, we use two different sets of assessments which the organizers create and NIST assessors.\footnote{NIST assessors are hired, trained, and compensated by the National Institute of Standards and Technology (NIST; \url{https://nist.gov}). }

During topic creation, the organizers annotated each turn in terms of their provenance to PTKB statements and included their labels in the released topic files. To ensure the quality of these annotations, we assigned each turn to at least two of the organizers. In cases of disagreement, we assigned the turns to a third annotator and assigned the majority vote label.
Moreover, during the assessment of passage relevance, the NIST assessors were also asked to judge the relevance of PTKB statements to each turn. The assessment pool in this case was smaller than the one done by the organizers. The organizers judged all of the turns, while the NIST assessors only judged the turns that were selected for passage relevance. 
As part of the iKAT resource, we provide both sets of assessments for this task.

\subsection{Assessments for Passage Retrieval Task}
\label{sec:passRet}
\begin{figure}
    \centering
    \includegraphics[width=\columnwidth]{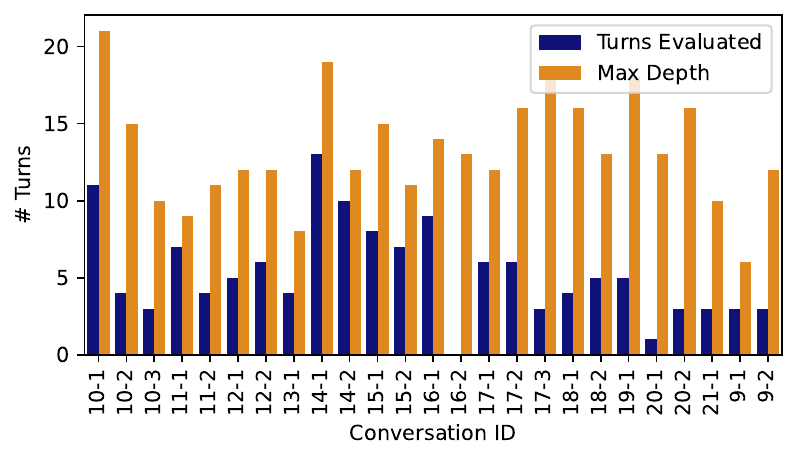}
    \caption{Number turns evaluated per dialogue in the final judgment pool vs.~the maximum depth of each topic.}
    \label{fig:turn-eval}
\end{figure}

The NIST assessors have judged the relevance of the passages based on the methodology used in CAsT (with the same scale). We selected a subset of 176 turns out of 326 to be judged by NIST assessors. Among the un-assessed turns, were responses that were clarifications (e.g., ``Do you have any dietary requirements?'') or were responses to utterances that were too general and returned too many relevant documents (e.g., ``I'm traveling to California, do you have any suggestions?'').
A pool of 26,159 passages was created and manually judged. An average number of 147 passages were judged for each turn. More detailed statistics of the collected data and judgments can be found in Table~\ref{tab:stats}. We also reported the number of turns per dialogue, as well as the number of turns evaluated per dialogue in Figure~\ref{fig:turn-eval}.

\begin{table}[]
    \centering
    \caption{Statistics of test data.}
    \begin{tabular}{ll}
    \toprule
        Topics & \phantom{00,0}12 \\
        Dialogues & \phantom{00,0}25 \\
        Turns & \phantom{00,}326 \\
        Assessed turns & \phantom{00,}176 \\ 
        Avg.~dialogue length & \phantom{00,0}13.04 \\
        Avg.~ num.~of dialogue per topic &  \phantom{00,00}1.92\\        
        Passages assessed & 26,159 \\
    \midrule
        Num of pruned turns  & \phantom{00,0}43 \\ 
        Num of turns after pruning  & \phantom{00,}133\\
        Num of dialogues after pruning  & \phantom{00,0}24 \\ 
    \midrule
        Fails to meet (0) & 20,458\\
        Slightly meets (1) & \phantom{0}2,787\\
        Moderately meets (2) & \phantom{0}1,803\\
        Highly meets (3) & \phantom{00,}932\\
        Fully meets (4) & \phantom{00,}179\\
    \midrule \midrule
        PTKB turns assessed by NIST & \phantom{00,0}98 \\
        PTKB assessments by NIST & \phantom{0}1,030 \\
        Relevant (1) & \phantom{00,}224 \\
    \midrule
        PTKB turns assessed by the organizers & \phantom{00,}112 \\
        PTKB assessments by the organizers & \phantom{0}1,158 \\
        Relevant (1) & \phantom{00,}182 \\
    \bottomrule            
    \end{tabular}    
    \label{tab:stats}
\end{table}

\subsection{Assessments for Response Generation Task - Human Judgements} \label{sec:human}
To evaluate the generated responses, we design a crowdsourcing task on Prolific\footnote{\url{https://prolific.com}} where we instruct the annotators to read the conversation between a system and a user and assess the quality of the last system response based on two aspects, namely, relevance and completeness. We ask the annotators to provide a graded label for each aspect. We provide a clear definition of the aspects, as well as each of the graded labels, as summarized below:

\textbf{Relevance:} Does the response follow on from previous utterances?
\begin{itemize}%
    \item \textit{-1: Unable to Judge} - Cannot determine the relevance of the response due to lack of context or other reasons.
    \item \textit{0: No} - Does not follow on from the previous utterances, seems to be completely random to the current conversation, seems to be a completely different conversation.
    \item \textit{1: Partially Relevant} - The response is partially off-topic; may be vaguely related, but too divergent from the conversation.
    \item \textit{2: Relevant} - Follows on, but it is not entirely clear why the response is being presented.
    \item \textit{3: Highly Relevant} - Directly follows on, and it is clear why the response is being presented.
\end{itemize}

\textbf{Completeness:} Does the response provide a thorough and comprehensive answer to the question posed?
\begin{itemize}%
    \item \textit{-1: Unable to Judge} - Cannot determine the completeness of the response due to lack of context or other reasons.
    \item \textit{0: No} - The response does not address the question at all or provides entirely unrelated information.
    \item \textit{1: Somewhat} - The response touches on the topic but misses significant details or only addresses one aspect of a multi-part question.
    \item \textit{2: Yes (but not completely)} - The response covers most aspects of the question but may miss minor details or nuances.
    \item \textit{3: Yes} - The response comprehensively addresses the question, providing a detailed and thorough answer that leaves no aspect untouched.
\end{itemize}
We recruited the Prolific workers through an open call on the website and compensated the workers according to their country's minimum wage.

\subsection{Assessments for Response Generation Task - GPT-4 Judgements} \label{sec:gpt4}
We prompt GPT-4 to assess the quality of responses. To do this, 
we select a subset of the turns for assessment, discarding generic turns, while preserving personalized turns. We assess the top one responses generated for each turn for each submission. We also screen the responses and filter out the low-quality responses. For example, if the response is not semantically similar to the top-ranked passages, or if it includes repeated sentences.

Given the subset of turns, we then select the passages participants indicated that they used to generate the response. In case they did not include the list of used passages, we consider the top 5 passages, as instructed in the guidelines. 
To avoid excessive and unnecessary assessment, we only consider the automatic top run of the teams, in case they were more effective than the median performance. 

We evaluate each response from two perspectives: groundedness and naturalness. The criteria and the definitions we provided to GPT-4 for the assessment of each are as follows:

\textbf{Naturalness:} Does the response sound human-like?
\begin{itemize}%
    \item \textit{0. No} - The response does not sound like something a human would say given the conversation.
    \item \textit{1. Somewhat} - Parts of the response can be generated by human, but it is overall not fluent.
    \item \textit{2. Slightly natural} - The response is almost human-like. The response is well-formed but is not natural.
    \item \textit{3. Yes (but not completely)} - The response is almost human-like. The response is well-formed and natural in most parts but has some parts that are not natural.
    \item \textit{4. Yes} - The response is perfectly human-like and fluent.
\end{itemize}

\textbf{Groundedness:} Does the response appropriately reference or connect to the information provided in the provenance passages? 

\begin{itemize}%
    \item \textit{0: No} - The response does not reference the information provided in the provenance passages or is entirely disconnected from it.
    \item \textit{1: Yes} - The response is directly based on the information provided in the provenance passages, accurately reflects this information, and utilizes it to enhance the response's relevance and completeness.
\end{itemize}

For naturalness, only the response is provided with the instructions, while for groundedness, the response and the provenance passages are included.
To ensure the quality of the assessments, we test multiple setups and prompts and compare them to a subset of responses that are manually labeled by the organizers.  We use the configuration that has the highest agreement with the labeled data. The final prompt used for the released set of labels is also provided in the repository enabling users of the resource to obtain similar judgments.

\begin{table*}[t]
\centering
\caption{Automatic evaluation of passage retrieval results. \GRG run names are highlighted with \textit{italic} font. Evaluation at retrieval cutoff of 1000.  The superscripts indicate run IDs that are significantly different, as determined by a two-sided paired t-test with a Bonferroni correction, at a significance level of $p < 0.05$. Given the space limit, we run a two-tailed paired t-test only on the labeled runs (a--i). 
}
 \label{tab:automatic-results}
  \resizebox{\textwidth}{!}{%
    \begin{tabular}{lllllllll}
\toprule
                                                  Run ID &  nDCG@3 &  nDCG@5 &   nDCG &   P@20 &  Recall@20 &  Recall &    mAP \\
\midrule
                                      (a) \textit{run-4-GPT-4} &  0.4382$^{\text{(\textit{all})}}$ &  0.4396$^{\text{(\textit{all})}}$ & 0.3479$^{\text{(\textit{all})}}$ & 0.3444$^{\text{(\textit{all})}}$ &     0.1821$^{\text{(\textit{all})}}$ &  0.3456$^{\text{(bscdfghi)}}$ & 0.1759$^{\text{(\textit{all})}}$ \\     
                    (b) \textit{georgetown\_infosense\_ikat\_run\_3} &  0.3083$^{\text{({adfghi})}}$ &  0.3109$^{\text{({adfghi})}}$ & 0.2097$^{\text{({adhi})}}$ & 0.2519$^{\text{({adfghi})}}$ &     0.1168$^{\text{({adfghi})}}$ &  0.1862$^{\text{({acdehi})}}$ & 0.1042$^{\text{({adghi})}}$ \\
                      (c) \textit{georgetown\_infosense\_ikat\_run\_2} &  0.2912$^{\text{(aghi)}}$ &  0.2955$^{\text{(afghi)}}$ & 0.2119$^{\text{(ahi)}}$ & 0.2643$^{\text{(afghi)}}$ &     0.1211$^{\text{(afghi)}}$ &  0.1862$^{\text{(adehi)}}$ & 0.1072$^{\text{(aghi)}}$ \\
                       (d) \textit{georgetown\_infosense\_ikat\_run\_1} &  0.2292$^{\text{(ab)}}$ &  0.2299$^{\text{(ab)}}$ & 0.1689$^{\text{(abhi)}}$ & 0.2109$^{\text{(abhi)}}$ &     0.1015$^{\text{(abgi)}}$ &  0.1613$^{\text{(abehi)}}$ & 0.0868$^{\text{(abhi)}}$ \\
                               (e) run\_automatic\_dense\_monot5 &  0.2167$^{\text{(a)}}$ &  0.2206$^{\text{(a)}}$ & 0.2147$^{\text{(aghi)}}$ & 0.1831$^{\text{(a)}}$ &     0.0812$^{\text{(a)}}$ &  0.3058$^{\text{(bcdfghi)}}$ & 0.0754$^{\text{(ahi)}}$ \\
                                                   (f) ConvGQR &  0.1652$^{\text{(ad)}}$ &  0.1623$^{\text{(acd)}}$ & 0.1518$^{\text{(ai)}}$ & 0.1421$^{\text{(acd)}}$ &     0.0611$^{\text{(acd)}}$ &  0.2034$^{\text{(aehi)}}$ & 0.0551$^{\text{(a)}}$ \\
               (g) run\_automatic\_dense\_damo\_canard\_16000\_recall &  0.1648$^{\text{(acd)}}$ &  0.1619$^{\text{(acd)}}$ & 0.1352$^{\text{(aei)}}$ & 0.1402$^{\text{(acd)}}$ &     0.0557$^{\text{(abcd)}}$ &  0.1664$^{\text{(aehi)}}$ & 0.0505$^{\text{(acd)}}$ \\
                              (h) uot-yahoo\_run\_llmnoptkb &  0.1433$^{\text{(acd)}}$ &  0.1469$^{\text{(acd)}}$ & 0.0759$^{\text{(abcde)}}$ & 0.1071$^{\text{(abcd)}}$ &     0.0525$^{\text{(acd)}}$ &  0.0525$^{\text{(abcdefg)}}$ & 0.0350$^{\text{(abcde)}}$ \\
                                (i)   llama2\_only\_10\_docs &  0.1389$^{\text{(acd)}}$ &  0.1466$^{\text{(acd)}}$ & 0.0756$^{\text{(abcdefg)}}$ & 0.1192$^{\text{(abcd)}}$ &     0.0553$^{\text{(abcd)}}$ &  0.0553$^{\text{(abcdefg)}}$ & 0.0376$^{\text{(abcde)}}$ \\
                           \phantom{(a)} run-1-llama-zero-shot &  0.1494 &  0.1437 & 0.0815 & 0.1165 &     0.0507 &  0.0742 & 0.0387 \\
                                    \phantom{(a)} run\_automatic\_llm\_damo &  0.1343 &  0.1411 & 0.1105 & 0.1102 &     0.0487 &  0.1401 & 0.0376 \\
                                               \phantom{(a)}  LLMConvGQR &  0.1318 &  0.1338 & 0.1200 & 0.1169 &     0.0523 &  0.1620 & 0.0461 \\
                                          \phantom{(a)}        cfda1 &  0.1323 &  0.1291 & 0.0941 & 0.1267 &     0.0536 &  0.0963 & 0.0444 \\
                                         \phantom{(a)}        cfda2 &  0.1282 &  0.1260 & 0.0916 & 0.1218 &     0.0510 &  0.0963 & 0.0421 \\
                         \phantom{(a)}     uot-yahoo\_run\_rankgpt35 &  0.1130 &  0.1070 & 0.0496 & 0.0801 &     0.0322 &  0.0322 & 0.0224 \\
                            \phantom{(a)}     uot-yahoo\_run\_monot5 &  0.1107 &  0.1062 & 0.0499 & 0.0823 &     0.0330 &  0.0330 & 0.0223 \\
                                     \phantom{(a)}   uot-yahoo\_run &  0.1086 &  0.1049 & 0.0495 & 0.0823 &     0.0330 &  0.0330 & 0.0222 \\
                   \phantom{(a)}   run\_automatic\_dense\_mini\_LM\_reranker &  0.1056 &  0.1047 & 0.0548 & 0.0812 &     0.0308 &  0.0496 & 0.0206 \\
                  \phantom{(a)}       run-2-llama-fine-tuned &  0.0826 &  0.0816 & 0.0457 & 0.0684 &     0.0301 &  0.0425 & 0.0202 \\
                                           \phantom{(a)}      cfda4 &  0.0836 &  0.0806 & 0.0759 & 0.0793 &     0.0362 &  0.0963 & 0.0311 \\
                                           \phantom{(a)}      cfda3 &  0.0836 &  0.0806 & 0.0759 & 0.0793 &     0.0362 &  0.0963 & 0.0311 \\     
                      \phantom{(a)}  GRILL\_Colbert\_BART2Summariser &  0.0667 &  0.0641 & 0.0558 & 0.0451 &     0.0278 &  0.0669 & 0.0184 \\
       \phantom{(a)} GRILL\_BM25\_T5Rewriter\_T5Ranker\_BARTSummariser &  0.0630 &  0.0620 & 0.0496 & 0.0579 &     0.0214 &  0.0636 & 0.0168 \\
     \phantom{(a)} GRILL\_BM25\_T5Rewriter\_T5Ranker\_BARTSummariser\_10 &  0.0572 &  0.0581 & 0.0356 & 0.0500 &     0.0224 &  0.0284 & 0.0172 \\
               \phantom{(a)} bm25\_rm3-auto-ptkb\_3-k\_100-num\_psg-3 &  0.0396 &  0.0450 & 0.0277 & 0.0429 &     0.0176 &  0.0257 & 0.0118 \\
\bottomrule
\end{tabular}}

\end{table*}

\subsection{Runs}
\label{sec:baselines}
We provide the runs as a resource, so researchers can use them for comparison. The runs include four baseline runs developed by the authors (two automatic and two manual runs) along with 24 runs from teams that participated in TREC iKAT 2023.
Most of the runs use LLMs in their pipelines and we observe two main pipelines, namely, retrieve then generate (\RG) and generate, retrieve, then generate (\GRG).
Most teams use a multi-step \RG pipeline consisting of the following: 
(1) PTKB statement relevance prediction;
(2) conversational rewriting (most incorporating the previous canonical responses as well as predicted relevance PTKB statements) and conversational query expansion;
(3) retrieval using traditional or dense IR model; and 
(4) multi-stage passage re-ranking with neural language models fine-tuned for point-wise (mono) and pairwise (duo) ranking. 
Below we list the details of baselines and a brief summary of each of the submitted runs.

\begin{itemize}[leftmargin=*]
    \item \textbf{bm25\_rm3-manual-ptkb\_3-k\_100-num\_psg-3}. We used BM25 +RM3, with the default configuration in Pyserini, to retrieve an initial set of 100 passages for each query. To refine the query, we manually selected the top 3 most relevant PTKB statements and appended them to the manually resolved query. With our rewritten query, we conducted a second round of retrieval using the standard BM25 method in Pyserini. This process also retrieved 100 passages. From this secondary set of 100 passages, we selected the top 3 based on their relevance. These selected passages were then used to construct a final response. For this task, we used a T5 model that has been fine-tuned on the News Summary dataset, available on HuggingFace.\footnote{\url{https://huggingface.co/mrm8488/t5-base-finetuned-summarize-news}}

    \item \textbf{bm25\_rm3-auto-ptkb\_3-k\_100-num\_psg-3}. This approach is analogous to the textit{bm25\_rm3-manual-ptkb\_3-k\_100-num\_psg-3} method but employs an automated process for query rewriting and PTKB statement selection.  Specifically, we rewrote the query automatically using a T5 model fine-tuned on the Canard dataset,\footnote{\url{https://huggingface.co/castorini/t5-base-canard}} and obtained relevant PTKB statements automatically by re-ranking the statements using \texttt{SentenceTransformers}.\footnote{\url{https://huggingface.co/cross-encoder/ms-marco-MiniLM-L-6-v2}}

    \item \textbf{llama2\_only\_10\_docs}. This pipeline executed several interactions with a LLaMA-2 7B model, each employing distinct prompts tailored for specific tasks. The initial call involved revising the most recent part of the ongoing conversation. The prompt, which included the entire conversation up to that point, was designed to guide the model in reformulating the latest utterance. This step aimed to optimize the utterance for more effective search results in subsequent steps. Following the rewrite, the next step involved evaluating the relevance of documents retrieved based on the revised utterance. In this phase, the prompt fed to the model included both the conversation (as updated from the first call) and a specific document. The model's task was to assess and score the document's relevance in relation to the conversation's context. We only ranked the top 10 documents in the interest of time.  The final call in the pipeline focused on generating an appropriate response to fulfill the user's information need. The prompt for this stage incorporated the top three documents identified as relevant from the previous step, along with the entire conversation. The model uses this information to craft a response that aligns with the user's query and the context provided by the conversation and the selected documents.

    \item \textbf{ColBERT\_llama2summariser\_manual} employed ColBERT for retrieval with manual queries, and LLama-2 7B for summarizing the top-3 passages.

    \item \textbf{uot-yahoo\_run\_llmnoptkb}. Passage retrieval for each utterance turn is conducted using Pyserini's \texttt{LuceneSearcher}. For re-ranking, multiple LLMs are used to re-rank the top five passages retrieved in each turn by pair-wise ranking. An aggregation of these results leads to a final ranking. Notably, both passage retrieval and re-ranking stages do not consider relevant PTKB statements and rely solely on rewritten utterances in each turn. This process also follows a zero-shot learning approach.

    \item \textbf{georgetown\_infosense\_ikat\_run\_1/2/3} These three runs utilize LLaMA to initially generate responses to user queries by integrating relevant PTKBs. In runs 1 and 2, passages deemed reliable through a TF-IDF and logarithmic regression model analysis are further processed. Run 3 distinguishes itself by selecting top passages with BM25 scoring instead. Across the runs, the FastChat T5 model~\citep{zheng2023judging} is employed to condense these passages into concise one or two-sentence summaries. These summaries are then ranked by relevance to the query and concatenated to formulate the final response.

    \item \textbf{ConvGQR}. Combines query rewriting and query expansion to train on the QReCC dataset then applied to the iKAT dataset. This run does not provide generated responses (only includes passage ranking).

    \item \textbf{LLMConvGQR}. Merges query rewriting and query expansion based on ChatGPT, applying query reformulation directly on the iKAT dataset.

    \item \textbf{run\_automatic\_dense\_monot5}. The process here also unfolds in two distinct steps. The first step encompasses dense retrieval of passages, followed by their re-ranking. In this method, automatic queries undergo rewriting through a custom-trained module based on BART, with fine-tuning conducted using the Samsum and Canard datasets. The re-ranking phase employs a T5-based model.

    \item \textbf{run-4-GPT-4}. In this run, the GPT-4 model initially generates an answer for each turn. Subsequently, GPT-4 is employed to produce five queries for each answer. These generated queries are then processed by a BM25 model and a cross-encoder \texttt{MiniLM} for re-ranking. The first two documents retrieved for each query are selected and supplied to GPT-4, which then generates the response text~\cite{abbasiantaeb2024generate}.

    \item \textbf{run-1-llama-zero-shot}. Query understanding and response generation in this run are based on zero-shot prompting of the LLaMa model, with no training data used for these tasks. Re-ranking is conducted using the cross-encoder model from HuggingFace, specifically the \texttt{ms-marco-MiniLM-L-12-v2} model, which is trained for passage ranking on the MS Marco dataset. This approach applies zero-shot prompting with the LLaMa 7B model for both response generation and query rewriting. \texttt{SentenceTransformers} is utilized for PTKB selection. For re-ranking, the cross-encoder model from HuggingFace (\texttt{ms-marco-MiniLM-L-12-v2}), trained on the MS Marco dataset, is used.

    \item \textbf{cfda1}. The datasets used include QReCC and CAsT. Dense retrieval model trained on the MS MARCO passage ranking collection. The retrieval process involves sparse retrieval, where re-ranking is performed by dense retrievers. Generative QA models are used for response generation.
\end{itemize}

\begin{figure*}
    \centering
    \includegraphics[scale=0.5]
    {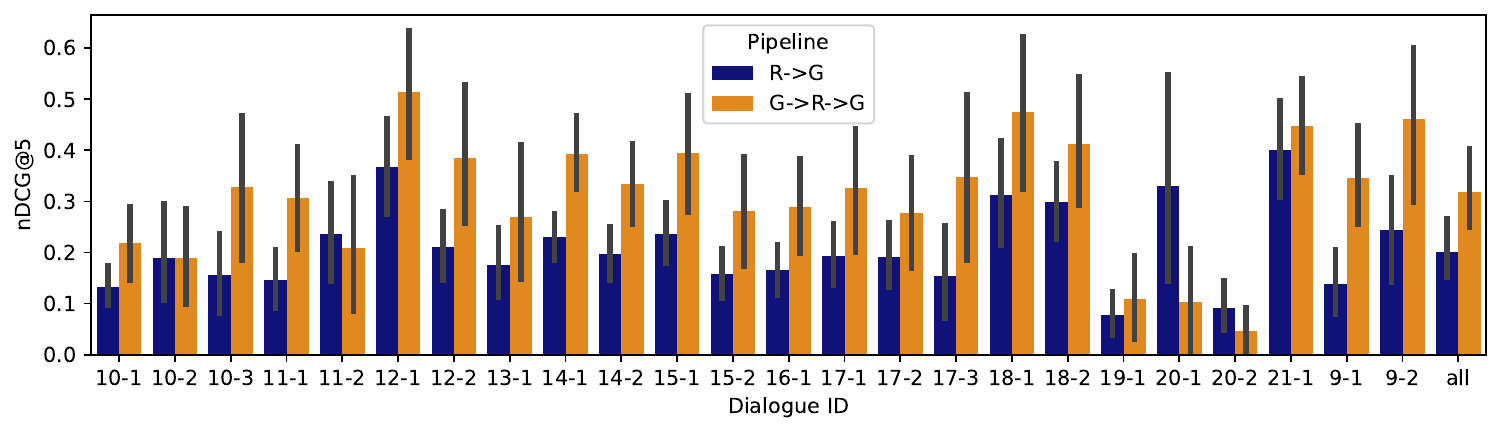}
    \caption{nDCG@5 aggregated for each dialogue across all runs on the passage ranking task. We report the average across runs, median or better.}
    \label{fig:dialogue-performance}
    \vspace{-2mm}
\end{figure*}

\begin{table*}
\caption{Performance of automatic runs on the PTKB provenance task based on NIST assessment. \GRG run names are highlighted with \textit{italic} font. The superscripts indicate run IDs that are significantly different, as determined by a two-sided paired t-test with a Bonferroni correction, at a significance level of $p < 0.05$. }
 \label{tab:ptkb-results}
    \begin{tabular}{llllll}
\toprule
                                          Run ID &  nDCG@3 &    P@3 &  Recall@3 &    MRR \\
\midrule
                            (a) run-1-llama-zero-shot &  0.7254$^{\text{(fhijk)}}$ & 0.4626$^{\text{(fghijk)}}$ &    0.6964$^{\text{(fghijk)}}$ & 0.7950$^{\text{(jk)}}$ \\
                           (b) run-2-llama-fine-tuned &  0.7102$^{\text{(fghijk)}}$ & 0.4490$^{\text{(fhijk)}}$ &    0.6796$^{\text{(fghijk)}}$ & 0.7795$^{\text{(fghijk)}}$ \\
                                   (c) uot-yahoo\_run &  0.6594$^{\text{(fghijk)}}$ & 0.4184$^{\text{(fghijk)}}$ &    0.6213$^{\text{(fghijk)}}$ & 0.7112$^{\text{(fghijk)}}$ \\
                                      (d) \textit{run-4-GPT-4} &  0.6174$^{\text{(fhijk)}}$ & 0.3605$^{\text{(hijk)}}$ &    0.5833$^{\text{(fhijk)}}$ & 0.7027$^{\text{(fjk)}}$ \\
                  (e) \textit{georgetown\_infosense\_ikat\_run\_1--3} &  0.4515$^{\text{(hijk)}}$ & 0.2551 &    0.4133$^{\text{(hij)}}$ & 0.5446$^{\text{(jk)}}$ \\
                  (f)  GRILL\_Colbert\_BART2Summariser &  0.3727$^{\text{(abcd)}}$ & 0.2483$^{\text{(abc)}}$ &    0.3836$^{\text{(abcd)}}$ & 0.5038$^{\text{(bcd)}}$ \\
                 (g) bm25\_rm3-auto-ptkb\_3-k\_100-num\_psg-3 &  0.3434$^{\text{(bc)}}$ & 0.2687$^{\text{(ac)}}$ &    0.3099$^{\text{(abc)}}$ & 0.3844$^{\text{(bc)}}$ \\
                                          (h) ConvGQR &  0.2934$^{\text{(abcdei)}}$ & 0.2109$^{\text{(abcdi)}}$ &    0.2756$^{\text{(abcdei)}}$ & 0.4419$^{\text{(bci)}}$ \\
                                       (i) LLMConvGQR &  0.2934$^{\text{(abcdeh)}}$ & 0.2109$^{\text{(abcdh)}}$ &    0.2756$^{\text{(abcdeh)}}$ & 0.4419$^{\text{(bch)}}$ \\
 (j) GRILL\_BM25\_T5Rewriter\_T5Ranker\_BARTSummariser\_10 &  0.2605$^{\text{(abcde)}}$ & 0.2211$^{\text{(abcd)}}$ &    0.2964$^{\text{(abcde)}}$ & 0.3757$^{\text{(abcde)}}$ \\
    (k) GRILL\_BM25\_T5Rewriter\_T5Ranker\_BARTSummariser &  0.2507$^{\text{(abcde)}}$ & 0.2075$^{\text{(abcd)}}$ &    0.3016$^{\text{(abcd)}}$ & 0.3756$^{\text{(abcde)}}$ \\
\bottomrule
\end{tabular}%

\end{table*}

\subsection{Publicly Available Resources}
To facilitate research in the area we have made the following resources publicly available on our GitHub repository:\footnote{\url{https://github.com/irlabamsterdam/iKAT}}
\begin{itemize}[nosep, leftmargin=*]
    \item The training and test topics.
    \item Python scripts that are used to segment the passages, segmented passages along with MD5 hashes, and the Pyserini index of the collection.
    \item PTKB relevance assessments judged by the organizers.
    \item PTKB and passage relevance assessments judged by NIST.
    \item ClueWeb22-b iKAT subset as well as the Pyserini index (available upon request, conditioned on obtaining ClueWeb22 license.\footnote{See \url{https://www.lemurproject.org/clueweb22/obtain.php}})
    \item Human- and GPT-4-generated quality labels of the top runs. 
    \item Prompt for generating GPT-4 labels.
    \item Baselines and participants' runs\footnote{Conditioned on the agreement of the participants.} described Section~\ref{sec:baselines}.
\end{itemize}

\section{Evaluation}
\label{sec:eval}

\subsubsection*{\bf Statement Ranking Task} We evaluate the PTKB statement ranking task at the turn and conversation levels. The ranking metrics include nDCG@3, P@3, Recall@5, and MRR.

\subsubsection*{\bf Passage Ranking Task.} In the main task, we assess the submissions across two key dimensions: ranking depth and turn depth. Ranking depth focuses on the earlier positions, specifically 3 and 5, which are critical in a conversational scenario where the top \(k\) results are utilized to generate responses. Turn depth examines performance at the \(n\)-th conversational turn, with higher performance in later turns indicating a stronger grasp of the preceding context.
The primary metric used for evaluation is mean nDCG@5, calculated by averaging scores across all conversational turns with uniform weights. We also detail turn-depth performance using nDCG@5, where scores for each turn are averaged at depth (\(n\)). Additional metrics include nDCG@3, P@10, Recall@10, and mean Average Precision (mAP), averaged across all turns for a comprehensive analysis of retrieval effectiveness.

\subsubsection*{\bf Response Generation Task.} Given the high likelihood of LLMs being used in this year's submissions and the possibility of hallucination, we evaluated the generated responses in terms of groundedness. Groundedness measures whether the generated response can be attributed to the passages that it is supposed to be generated from. We use GPT-4 to evaluate the relevance, completeness, groundedness, and naturalness of the responses, as it demonstrated a high correlation with human labels in our preliminary experiments. For each turn, we used the GPT-4 assessments and took the mean for each metric over all turns.

\section{Results \& Analysis}

To provide an overview of the challenges iKAT poses to the current state-of-the-art IR systems we report the results of the participants' runs, as well as organizers' baselines. The systems are evaluated on their ability to model the context and user personal knowledge, passage ranking, as well as response generation. We summarize the results of both \RG and \GRG methods, providing a comparison on how different retrieval and generation approaches benefit the systems.

\subsection{Passage Ranking}

\subsubsection{Overall results.}
Table~\ref{tab:automatic-results} lists the performance of the automatic runs in terms of all the evaluation metrics. We see that \GRG runs tend to perform better than \RG runs, suggesting that leveraging the learned knowledge of LLMs (GPT-4 and Llama in this case) leads to a better starting point for subsequent retrieval of relevant results and then the generation of a relevant response.

\subsubsection{Performance per dialogue.} 
Figure~\ref{fig:dialogue-performance} reports the average performance in terms of nDCG@5 of all runs that have a median or better performance. 
We see that while the runs perform well for some of the topics, they fail to perform well for some. In particular, we find topics 12-1 and 21-1 to be the easiest, while 19-1 and 20-2 to be the most difficult.

\subsubsection{Performance at different depths.} 
\begin{figure}
    \centering
    \includegraphics[width=.97\columnwidth]{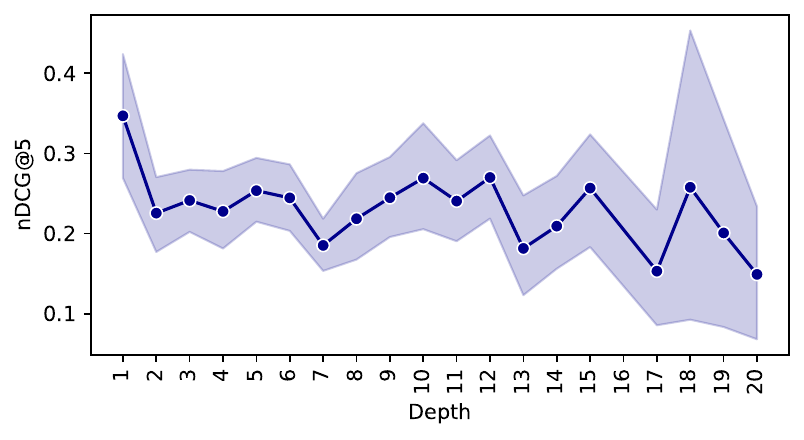}
    \caption{nDCG@5 at varying conversation turn depths on the passage ranking task. We report the average across runs, median or better.}
    \label{fig:turn-performance}
\end{figure}

\begin{figure}
    \centering
    \includegraphics[width=\columnwidth]{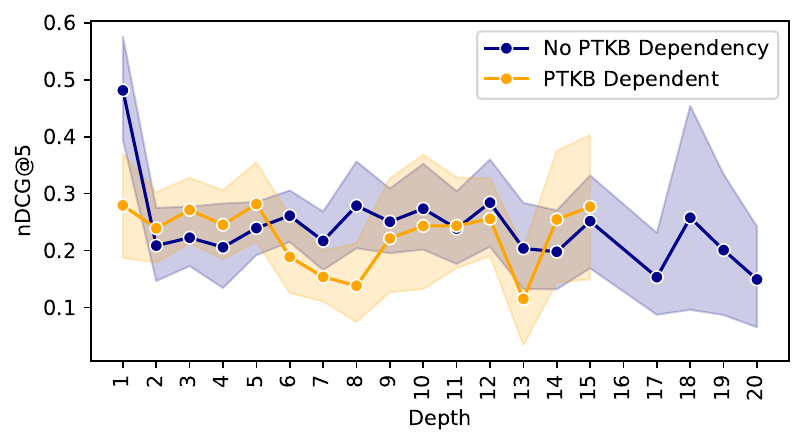}
    \caption{nDCG@5 at varying conversation turn depths on the passage ranking task, for turns that depend on PTKB statements vs.~those that do not. We report the average across runs, median or better.}
    \label{fig:turn-performance-ptkb}
\end{figure}

Figure~\ref{fig:turn-performance} reports the performance of all runs (median or better) at varying conversation turns in terms of nDCG@5. We also report the performance at different depths, separating the turns that depend on PTKB provenance in Figure~\ref{fig:turn-performance-ptkb}. Our intuition is that the PTKB statement ranking step will introduce additional difficulty and error in the pipeline and consequently the runs exhibit lower performance. However, we see that this is not always the case, and in most cases, PTKB dependence leads to lower performance. 
Similar to CAsT, we see that the models perform best in the first turn, and as the conversation progresses the performance becomes lower, with some peaks in the middle of the conversation.
As we compare the performance of the turns based on PTKB dependence, interestingly we see the highest difference in the first turn, suggesting that the significance of predicting the right PTKB statements in the early turns is essential and the task is more difficult in those earlier turns.

\subsection{PTKB Provenance}

\subsubsection{Overall results.} 
As previously described, we evaluated the submissions for the PTKB statement ranking task based on two relevance judgments, namely, assessed by the NIST assessors, as well as the organizers. We report the results based on NIST assessments in Table~\ref{tab:ptkb-results}. We see that \GRG models are not the top runs, despite their success in passage ranking, suggesting that while the LLMs can leverage PTKB statements effectively in response generation, they are not as effective in ranking the relevant PTKB statements in the \GRG pipeline. 
LLaMA in the zero-shot setting, however, achieved the best result in the PTKB statement ranking task based on both results.

\subsubsection{Performance per dialogue.}
\begin{figure}[!b]
    \centering
    \includegraphics[width=\columnwidth]{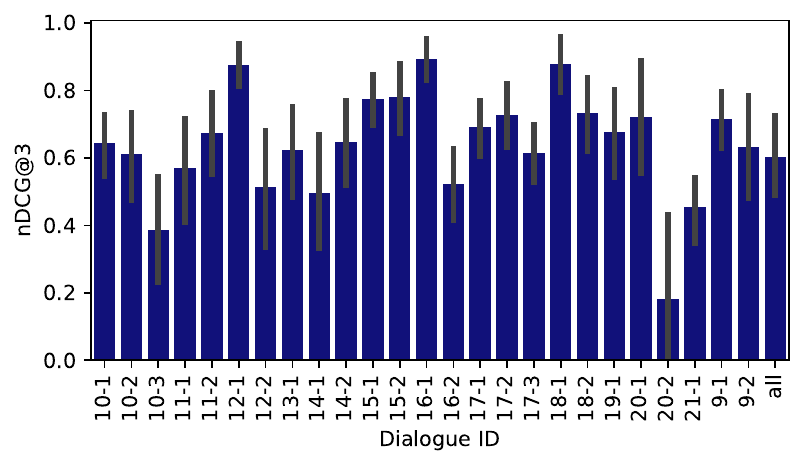}
    \vspace{-2mm}
    \caption{nDCG@3 on PTKB relevance prediction, aggregated for each topic across all runs. We report the average across runs, median or better.}
    \label{fig:ptkb-dialogue}
\end{figure}

Using the organizers' assessments, in Figure~\ref{fig:ptkb-dialogue} we plotted the mean performance of all the submissions (median and better) in terms of nDCG@3, aggregated on each topic. While we observed a reasonably high performance for all the topics, we find topic 20-2 to be the most challenging for this task, and 16-1 to be the easiest one. Comparing the results of this table with Table~\ref{fig:dialogue-performance}, surprisingly we do not notice a clear correlation between PTKB statement ranking and passage retrieval performance.

\begin{table}[!b]
    \centering
    \caption{Evaluating the Relevance and Completeness of the responses by human assessors. The superscripts indicate run IDs that are significantly different, as determined by a two-sided paired t-test with a Bonferroni correction, at a significance level of $p < 0.05$.}
    \resizebox{\columnwidth}{!}{
    \begin{tabular}{llll}
    \toprule
          Run     & Relevance & Completeness \\ 
    \midrule
         (a) run-4-GPT-4                           & 2.42$^{\text{(bcde)}}$ & 2.54$^{\text{(bcde)}}$ \\ 
         (b) georgetown\_infosense\_ikat\_run\_3   & 1.78$^{\text{(ac)}}$ & 1.90$^{\text{(acd)}}$ \\
         (c) uot-yahoo\_run\_llmnoptkb             & 1.18$^{\text{(abe)}}$ & 1.11$^{\text{(abe)}}$ \\
         (d) run\_automatic\_dense\_monot5         & 1.31$^{\text{(a)}}$ &  1.24$^{\text{(a)}}$ \\
         (e) llama2\_only\_10\_docs                & 1.63$^{\text{(ac)}}$ & 1.65$^{\text{(ae)}}$ \\
    \bottomrule
    \end{tabular}
    }
    \label{tab:human}
\end{table}

\begin{table*}[]
    \centering
    \caption{Evaluating the Relevance, Completeness, Groundedness, and Naturalness of the responses using GPT-4 as external evaluator.  The superscripts indicate run IDs that are significantly different, as determined by a two-sided paired t-test with a Bonferroni correction, at a significance level of $p < 0.05$.}
    \vspace{-2mm}
    \begin{tabular}{llllll}
    \toprule
         Run                                  & Relevance & Completeness & Groundedness & Naturalness \\ 
    \midrule
        (a) run-4-GPT-4                          & 2.92$^{\text{(bcde)}}$ & 2.85$^{\text{(bcde)}}$ & 0.89 (65/73)$^{\text{(bcde)}}$ & 4.00$^{\text{(bcde)}}$ \\ 
        (b) georgetown\_infosense\_ikat\_run\_3  & 1.56$^{\text{(acd)}}$    & 1.23$^{\text{(acd)}}$ & 0.68 (50/73)$^{\text{(ae)}}$ & 3.64$^{\text{(acd)}}$ \\
        (c) uot-yahoo\_run\_llmnoptkb            & 0.35$^{\text{(abe)}}$ & 0.11$^{\text{(abe)}}$ & 0.67 (49/73)$^{\text{(ae)}}$ & 2.90$^{\text{(abe)}}$ \\       
        (d) run\_automatic\_dense\_monot5        & 0.69$^{\text{(abe)}}$ & 0.37$^{\text{(abe)}}$ & 0.51 (37/73)$^{\text{(a)}}$ & 2.77$^{\text{(abe)}}$ \\
        (e) llama2\_only\_10\_docs               & 1.26$^{\text{(acd)}}$ & 1.03$^{\text{(acd)}}$ & 0.31 (23/73)$^{\text{(abc)}}$ & 3.70$^{\text{(acd)}}$ \\
    \bottomrule
    \end{tabular}%
    \label{tab:groundedness}
    \vspace{-3mm}
\end{table*}

\subsection{Response Evaluation}
We evaluate the quality of the generated responses from the top runs of the five best-performing teams, selecting the,e highest-ranking submission from each team (above the median) and the best baseline model for comparison. It is important to note that the top submission from \texttt{RALI} (\texttt{ConvGQR}) does not include a generated response. Our evaluation focuses solely on the 73 turns that have been assessed by NIST and are included in the teams' final evaluations.

\subsubsection{Human evaluation}

Given the complexities of the annotation task and constraints on our budget, we limited our collection of human labels to relevance and completeness. As previously discussed, the labels for these aspects were graded. Table~\ref{tab:human} presents the average scores for relevance and completeness across all evaluated turns. Notably, \texttt{run-4-GPT-4} significantly outperforms all other methods in both relevance and completeness, mirroring its success in passage ranking. 

Following the trends in passage ranking results, \texttt{georgetown\_infosense\_ikat\_run\_3} ranks as the second-best method in both metrics; however, the improvements over some methods are not statistically significant. Surprisingly, the Organizers' baseline model, \texttt{llama2\_only\_10\_docs}, is ranked third, outperforming two other models. This ranking is unexpected, as it does not align with the passage ranking results shown in Table~\ref{tab:automatic-results}.
 
\subsubsection{GPT-4 evaluation}

In light of concerns about the hallucination and unfaithfulness of generated responses, we evaluated top runs for groundedness and other metrics, as shown in Table~\ref{tab:groundedness}. Additionally, we conducted a GPT-4 evaluation for relevance and completeness to compare with human assessments. The GPT-4-based model significantly outperformed other models in terms of relevance and completeness. Notably, the run by \texttt{InfoSense} was ranked second, and the Organizers' baseline ranked third.

In terms of naturalness, the baseline run was only marginally ranked second, a difference not considered statistically significant. In terms of groundedness, it is interesting to note that the model \texttt{llama2\_only\_10\_docs} ranked as the least effective, with only 23 out of 73 responses being deemed grounded according to the order in Table~\ref{tab:automatic-results}. This indicates that the high scores of this model might stem from its unfaithfulness to the retrieved passages, possibly relying on its internal knowledge to generate responses. Similar trends were observed with other methods across different metrics.

\subsubsection{Potential biases}
While comparing the results obtained from human and GPT-4 labels in Tables~\ref{tab:human} and \ref{tab:groundedness}, we observe the same ranking of the methods in terms of both relevance and completeness, indicating that we could rely on the GPT-4 labels to evaluate newly generated responses, even if GPT-4 itself is among the runs. However, we clearly see that the absolute pointwise differences between the performance of the models are different, where the GPT-4 run seems to have been favored by the GPT-4 as evaluator, in line with the findings of \citet{DBLP:journals/corr/abs-2311-09766}, showing that LLMs tend to favor the text generated by themselves when they act as evaluators. Nevertheless, we find that for naturalness and completeness, they can still be used to rank the models.

\section{Conclusion}

We introduce the iKAT resources, which build on the foundations established by TREC iKAT 2023. The iKAT resources allow researchers to assess conversational information seeking across various personas, distinguishing our test collection from others like SQuAD and CAsT. The unique aspect of our resource is its emphasis on handling personalized and complex conversations, which necessitate advanced reasoning and the effective use of personal knowledge graphs to generate relevant responses. Looking ahead, we aim to expand this resource to develop a more adaptable and scalable framework for evaluating personalized conversational agents across a broader array of topics and personas.

\subsubsection*{\bf Acknowledgments}
This material is based upon work partially supported by the Engineering and Physical Sciences Research Council (EPSRC) grant EP/V025708/1. Any opinions, findings, and conclusions or recommendations expressed in this material are those of the author(s) and do not necessarily reflect the views of the EPSRC.

We are immensely grateful for Jamie Callan's significant support and patience, which were instrumental in allowing us to provide a subset of the ClueWeb'22 collection to our participants. We also extend our gratitude to Andrew Ramsay, who assisted in hosting and delivering the collection to the participants.

Our thanks also go to Ian Soboroff for his extensive experience, patience, and persistence in managing the assessment process. %

\bibliographystyle{ACM-Reference-Format}
\bibliography{bibliography}

%%% -*-BibTeX-*-
%%% Do NOT edit. File created by BibTeX with style
%%% ACM-Reference-Format-Journals [18-Jan-2012].

\begin{thebibliography}{31}

%%% ====================================================================
%%% NOTE TO THE USER: you can override these defaults by providing
%%% customized versions of any of these macros before the \bibliography
%%% command.  Each of them MUST provide its own final punctuation,
%%% except for \shownote{}, \showDOI{}, and \showURL{}.  The latter two
%%% do not use final punctuation, in order to avoid confusing it with
%%% the Web address.
%%%
%%% To suppress output of a particular field, define its macro to expand
%%% to an empty string, or better, \unskip, like this:
%%%
%%% \newcommand{\showDOI}[1]{\unskip}   % LaTeX syntax
%%%
%%% \def \showDOI #1{\unskip}           % plain TeX syntax
%%%
%%% ====================================================================

\ifx \showCODEN    \undefined \def \showCODEN     #1{\unskip}     \fi
\ifx \showDOI      \undefined \def \showDOI       #1{#1}\fi
\ifx \showISBNx    \undefined \def \showISBNx     #1{\unskip}     \fi
\ifx \showISBNxiii \undefined \def \showISBNxiii  #1{\unskip}     \fi
\ifx \showISSN     \undefined \def \showISSN      #1{\unskip}     \fi
\ifx \showLCCN     \undefined \def \showLCCN      #1{\unskip}     \fi
\ifx \shownote     \undefined \def \shownote      #1{#1}          \fi
\ifx \showarticletitle \undefined \def \showarticletitle #1{#1}   \fi
\ifx \showURL      \undefined \def \showURL       {\relax}        \fi
% The following commands are used for tagged output and should be
% invisible to TeX
\providecommand\bibfield[2]{#2}
\providecommand\bibinfo[2]{#2}
\providecommand\natexlab[1]{#1}
\providecommand\showeprint[2][]{arXiv:#2}

\bibitem[Abbasiantaeb and Aliannejadi(2024)]%
        {abbasiantaeb2024generate}
\bibfield{author}{\bibinfo{person}{Zahra Abbasiantaeb} {and} \bibinfo{person}{Mohammad Aliannejadi}.} \bibinfo{year}{2024}\natexlab{}.
\newblock \bibinfo{title}{Generate then Retrieve: Conversational Response Retrieval Using LLMs as Answer and Query Generators}.
\newblock
\newblock
\showeprint[arxiv]{2403.19302}~[cs.IR]


\bibitem[Abbasiantaeb et~al\mbox{.}(2024)]%
        {DBLP:conf/wsdm/Abbasiantaeb0KA24}
\bibfield{author}{\bibinfo{person}{Zahra Abbasiantaeb}, \bibinfo{person}{Yifei Yuan}, \bibinfo{person}{Evangelos Kanoulas}, {and} \bibinfo{person}{Mohammad Aliannejadi}.} \bibinfo{year}{2024}\natexlab{}.
\newblock \showarticletitle{Let the LLMs Talk: Simulating Human-to-Human Conversational {QA} via Zero-Shot LLM-to-LLM Interactions}. In \bibinfo{booktitle}{\emph{International Conference on Web Search and Data Mining ({WSDM})}}. \bibinfo{publisher}{{ACM}}, \bibinfo{pages}{8--17}.
\newblock


\bibitem[Aliannejadi et~al\mbox{.}(2024)]%
        {aliannejadi2024trec}
\bibfield{author}{\bibinfo{person}{Mohammad Aliannejadi}, \bibinfo{person}{Zahra Abbasiantaeb}, \bibinfo{person}{Shubham Chatterjee}, \bibinfo{person}{Jeffery Dalton}, {and} \bibinfo{person}{Leif Azzopardi}.} \bibinfo{year}{2024}\natexlab{}.
\newblock \showarticletitle{TREC iKAT 2023: The Interactive Knowledge Assistance Track Overview}. In \bibinfo{booktitle}{\emph{Text REtrieval Conference ({TREC})}}. \bibinfo{publisher}{NIST}.
\newblock


\bibitem[Azzopardi et~al\mbox{.}(2018)]%
        {azzopardi2018conceptualizing}
\bibfield{author}{\bibinfo{person}{Leif Azzopardi}, \bibinfo{person}{Mateusz Dubiel}, \bibinfo{person}{Martin Halvey}, {and} \bibinfo{person}{Jeffery Dalton}.} \bibinfo{year}{2018}\natexlab{}.
\newblock \showarticletitle{Conceptualizing agent-human interactions during the conversational search process}. In \bibinfo{booktitle}{\emph{The second international workshop on conversational approaches to information retrieval}}.
\newblock


\bibitem[Belkin(2008)]%
        {belkin2008iir}
\bibfield{author}{\bibinfo{person}{Nicholas~J. Belkin}.} \bibinfo{year}{2008}\natexlab{}.
\newblock \showarticletitle{Some(what) grand challenges for information retrieval}.
\newblock \bibinfo{journal}{\emph{SIGIR Forum}} \bibinfo{volume}{42}, \bibinfo{number}{1} (\bibinfo{date}{jun} \bibinfo{year}{2008}), \bibinfo{pages}{47–54}.
\newblock
\showISSN{0163-5840}
\urldef\tempurl%
\url{https://doi.org/10.1145/1394251.1394261}
\showDOI{\tempurl}


\bibitem[Belkin et~al\mbox{.}(1996)]%
        {DBLP:conf/trec/BelkinCCKNPPRSX96}
\bibfield{author}{\bibinfo{person}{Nicholas~J. Belkin}, \bibinfo{person}{A. Cabezas}, \bibinfo{person}{Colleen Cool}, \bibinfo{person}{K. Kim}, \bibinfo{person}{Kwong~Bor Ng}, \bibinfo{person}{Soyeon Park}, \bibinfo{person}{R. Pressman}, \bibinfo{person}{Soo~Young Rieh}, \bibinfo{person}{Pamela~A. Savage{-}Knepshield}, {and} \bibinfo{person}{Hong~(Iris) Xie}.} \bibinfo{year}{1996}\natexlab{}.
\newblock \showarticletitle{Rutgers Interactive Track at {TREC-5}}. In \bibinfo{booktitle}{\emph{Text REtrieval Conference ({TREC})}}. \bibinfo{publisher}{NIST}.
\newblock


\bibitem[Dalton et~al\mbox{.}(2020a)]%
        {dalton2019cast}
\bibfield{author}{\bibinfo{person}{Jeffrey Dalton}, \bibinfo{person}{Chenyan Xiong}, {and} \bibinfo{person}{Jamie Callan}.} \bibinfo{year}{2020}\natexlab{a}.
\newblock \showarticletitle{{CAsT} 2019: The conversational assistance track overview}. In \bibinfo{booktitle}{\emph{Text REtrieval Conference ({TREC})}}. \bibinfo{publisher}{NIST}.
\newblock


\bibitem[Dalton et~al\mbox{.}(2021)]%
        {dalton2021cast}
\bibfield{author}{\bibinfo{person}{Jeffrey Dalton}, \bibinfo{person}{Chenyan Xiong}, {and} \bibinfo{person}{Jamie Callan}.} \bibinfo{year}{2021}\natexlab{}.
\newblock \showarticletitle{{TREC CAsT} 2021: The Conversational Assistance Track Overview}. In \bibinfo{booktitle}{\emph{Text REtrieval Conference ({TREC})}}. \bibinfo{publisher}{NIST}.
\newblock


\bibitem[Dalton et~al\mbox{.}(2020b)]%
        {dalton2020cast}
\bibfield{author}{\bibinfo{person}{Jeffrey Dalton}, \bibinfo{person}{Chenyan Xiong}, \bibinfo{person}{Vaibhav Kumar}, {and} \bibinfo{person}{Jamie Callan}.} \bibinfo{year}{2020}\natexlab{b}.
\newblock \showarticletitle{{CAsT-19}: {A} Dataset for Conversational Information Seeking}. In \bibinfo{booktitle}{\emph{International {ACM} {SIGIR} Conference on Research and Development in Information Retrieval}}. \bibinfo{publisher}{{ACM}}, \bibinfo{pages}{1985--1988}.
\newblock


\bibitem[Devlin et~al\mbox{.}(2019)]%
        {devlin2019bert}
\bibfield{author}{\bibinfo{person}{Jacob Devlin}, \bibinfo{person}{Ming-Wei Chang}, \bibinfo{person}{Kenton Lee}, {and} \bibinfo{person}{Kristina Toutanova}.} \bibinfo{year}{2019}\natexlab{}.
\newblock \bibinfo{title}{BERT: Pre-training of Deep Bidirectional Transformers for Language Understanding}.
\newblock
\newblock
\showeprint[arxiv]{1810.04805}~[cs.CL]


\bibitem[Guo et~al\mbox{.}(2023)]%
        {guo2023evaluating}
\bibfield{author}{\bibinfo{person}{Zishan Guo}, \bibinfo{person}{Renren Jin}, \bibinfo{person}{Chuang Liu}, \bibinfo{person}{Yufei Huang}, \bibinfo{person}{Dan Shi}, \bibinfo{person}{Supryadi}, \bibinfo{person}{Linhao Yu}, \bibinfo{person}{Yan Liu}, \bibinfo{person}{Jiaxuan Li}, \bibinfo{person}{Bojian Xiong}, {and} \bibinfo{person}{Deyi Xiong}.} \bibinfo{year}{2023}\natexlab{}.
\newblock \bibinfo{title}{Evaluating Large Language Models: A Comprehensive Survey}.
\newblock
\newblock
\showeprint[arxiv]{2310.19736}~[cs.CL]


\bibitem[Hersh(2002)]%
        {DBLP:conf/trec/Hersh02}
\bibfield{author}{\bibinfo{person}{William~R. Hersh}.} \bibinfo{year}{2002}\natexlab{}.
\newblock \showarticletitle{{TREC} 2002 Interactive Track Report}. In \bibinfo{booktitle}{\emph{Text REtrieval Conference ({TREC})}}, \bibfield{editor}{\bibinfo{person}{Ellen~M. Voorhees} {and} \bibinfo{person}{Lori~P. Buckland}} (Eds.). \bibinfo{publisher}{NIST}.
\newblock


\bibitem[Liang et~al\mbox{.}(2023)]%
        {liang2023holistic}
\bibfield{author}{\bibinfo{person}{Percy Liang}, \bibinfo{person}{Rishi Bommasani}, \bibinfo{person}{Tony Lee}, \bibinfo{person}{Dimitris Tsipras}, \bibinfo{person}{Dilara Soylu}, \bibinfo{person}{Michihiro Yasunaga}, \bibinfo{person}{Yian Zhang}, \bibinfo{person}{Deepak Narayanan}, \bibinfo{person}{Yuhuai Wu}, \bibinfo{person}{Ananya Kumar}, \bibinfo{person}{Benjamin Newman}, \bibinfo{person}{Binhang Yuan}, \bibinfo{person}{Bobby Yan}, \bibinfo{person}{Ce Zhang}, \bibinfo{person}{Christian Cosgrove}, \bibinfo{person}{Christopher~D. Manning}, \bibinfo{person}{Christopher Ré}, \bibinfo{person}{Diana Acosta-Navas}, \bibinfo{person}{Drew~A. Hudson}, \bibinfo{person}{Eric Zelikman}, \bibinfo{person}{Esin Durmus}, \bibinfo{person}{Faisal Ladhak}, \bibinfo{person}{Frieda Rong}, \bibinfo{person}{Hongyu Ren}, \bibinfo{person}{Huaxiu Yao}, \bibinfo{person}{Jue Wang}, \bibinfo{person}{Keshav Santhanam}, \bibinfo{person}{Laurel Orr}, \bibinfo{person}{Lucia Zheng}, \bibinfo{person}{Mert Yuksekgonul},
  \bibinfo{person}{Mirac Suzgun}, \bibinfo{person}{Nathan Kim}, \bibinfo{person}{Neel Guha}, \bibinfo{person}{Niladri Chatterji}, \bibinfo{person}{Omar Khattab}, \bibinfo{person}{Peter Henderson}, \bibinfo{person}{Qian Huang}, \bibinfo{person}{Ryan Chi}, \bibinfo{person}{Sang~Michael Xie}, \bibinfo{person}{Shibani Santurkar}, \bibinfo{person}{Surya Ganguli}, \bibinfo{person}{Tatsunori Hashimoto}, \bibinfo{person}{Thomas Icard}, \bibinfo{person}{Tianyi Zhang}, \bibinfo{person}{Vishrav Chaudhary}, \bibinfo{person}{William Wang}, \bibinfo{person}{Xuechen Li}, \bibinfo{person}{Yifan Mai}, \bibinfo{person}{Yuhui Zhang}, {and} \bibinfo{person}{Yuta Koreeda}.} \bibinfo{year}{2023}\natexlab{}.
\newblock \bibinfo{title}{Holistic Evaluation of Language Models}.
\newblock
\newblock
\showeprint[arxiv]{2211.09110}~[cs.CL]


\bibitem[Liu et~al\mbox{.}(2023)]%
        {DBLP:journals/corr/abs-2311-09766}
\bibfield{author}{\bibinfo{person}{Yiqi Liu}, \bibinfo{person}{Nafise~Sadat Moosavi}, {and} \bibinfo{person}{Chenghua Lin}.} \bibinfo{year}{2023}\natexlab{}.
\newblock \bibinfo{title}{LLMs as Narcissistic Evaluators: When Ego Inflates Evaluation Scores}.
\newblock
\newblock
\showeprint[arxiv]{2311.09766}~[cs.CL]


\bibitem[Over(2001)]%
        {over2001trec}
\bibfield{author}{\bibinfo{person}{Paul Over}.} \bibinfo{year}{2001}\natexlab{}.
\newblock \showarticletitle{The TREC interactive track: an annotated bibliography}.
\newblock \bibinfo{journal}{\emph{Information Processing \& Management}} \bibinfo{volume}{37}, \bibinfo{number}{3} (\bibinfo{year}{2001}), \bibinfo{pages}{369--381}.
\newblock


\bibitem[Overwijk et~al\mbox{.}(2022)]%
        {overwijk2022clueweb22}
\bibfield{author}{\bibinfo{person}{Arnold Overwijk}, \bibinfo{person}{Chenyan Xiong}, \bibinfo{person}{Xiao Liu}, \bibinfo{person}{Cameron VandenBerg}, {and} \bibinfo{person}{Jamie Callan}.} \bibinfo{year}{2022}\natexlab{}.
\newblock \bibinfo{title}{ClueWeb22: 10 Billion Web Documents with Visual and Semantic Information}.
\newblock
\newblock
\showeprint[arxiv]{2211.15848}~[cs.IR]


\bibitem[Owoicho et~al\mbox{.}(2023)]%
        {owoicho2023trec}
\bibfield{author}{\bibinfo{person}{Paul Owoicho}, \bibinfo{person}{Jeffrey Dalton}, \bibinfo{person}{Mohammad Aliannejadi}, \bibinfo{person}{Leif Azzopardi}, \bibinfo{person}{Johanne~R Trippas}, {and} \bibinfo{person}{Svitlana Vakulenko}.} \bibinfo{year}{2023}\natexlab{}.
\newblock \showarticletitle{TREC CAsT 2022: Going Beyond User Ask and System Retrieve with Initiative and Response Generation}. In \bibinfo{booktitle}{\emph{Text REtrieval Conference ({TREC})}}. \bibinfo{publisher}{NIST}.
\newblock


\bibitem[Radlinski and Craswell(2017)]%
        {radlinski2017theoretical}
\bibfield{author}{\bibinfo{person}{Filip Radlinski} {and} \bibinfo{person}{Nick Craswell}.} \bibinfo{year}{2017}\natexlab{}.
\newblock \showarticletitle{A theoretical framework for conversational search}. In \bibinfo{booktitle}{\emph{Proceedings of the 2017 Conference on Conference Human Information Interaction and Retrieval}}. ACM, \bibinfo{pages}{117--126}.
\newblock


\bibitem[Rahmani et~al\mbox{.}(2024)]%
        {DBLP:conf/eacl/RahmaniWANY24}
\bibfield{author}{\bibinfo{person}{Hossein~A. Rahmani}, \bibinfo{person}{Xi Wang}, \bibinfo{person}{Mohammad Aliannejadi}, \bibinfo{person}{Mohammadmehdi Naghiaei}, {and} \bibinfo{person}{Emine Yilmaz}.} \bibinfo{year}{2024}\natexlab{}.
\newblock \showarticletitle{Clarifying the Path to User Satisfaction: An Investigation into Clarification Usefulness}. In \bibinfo{booktitle}{\emph{Findings of the Association for Computational Linguistics ({EACL})}}. \bibinfo{publisher}{Association for Computational Linguistics}, \bibinfo{pages}{1266--1277}.
\newblock


\bibitem[Rajpurkar et~al\mbox{.}(2016)]%
        {rajpurkar2016squad}
\bibfield{author}{\bibinfo{person}{Pranav Rajpurkar}, \bibinfo{person}{Jian Zhang}, \bibinfo{person}{Konstantin Lopyrev}, {and} \bibinfo{person}{Percy Liang}.} \bibinfo{year}{2016}\natexlab{}.
\newblock \bibinfo{title}{SQuAD: 100,000+ Questions for Machine Comprehension of Text}.
\newblock
\newblock
\showeprint[arxiv]{1606.05250}~[cs.CL]


\bibitem[Reddy et~al\mbox{.}(2019)]%
        {reddy2019coqa}
\bibfield{author}{\bibinfo{person}{Siva Reddy}, \bibinfo{person}{Danqi Chen}, {and} \bibinfo{person}{Christopher~D. Manning}.} \bibinfo{year}{2019}\natexlab{}.
\newblock \bibinfo{title}{CoQA: A Conversational Question Answering Challenge}.
\newblock
\newblock
\showeprint[arxiv]{1808.07042}~[cs.CL]


\bibitem[Russell-Rose et~al\mbox{.}(2018)]%
        {russell2018information}
\bibfield{author}{\bibinfo{person}{Tony Russell-Rose}, \bibinfo{person}{Jon Chamberlain}, {and} \bibinfo{person}{Leif Azzopardi}.} \bibinfo{year}{2018}\natexlab{}.
\newblock \showarticletitle{Information retrieval in the workplace: A comparison of professional search practices}.
\newblock \bibinfo{journal}{\emph{Information Processing \& Management}} \bibinfo{volume}{54}, \bibinfo{number}{6} (\bibinfo{year}{2018}), \bibinfo{pages}{1042--1057}.
\newblock


\bibitem[Srivastava et~al\mbox{.}(2023)]%
        {srivastava2023imitation}
\bibfield{author}{\bibinfo{person}{Aarohi Srivastava}, \bibinfo{person}{Abhinav Rastogi}, \bibinfo{person}{Abhishek Rao}, \bibinfo{person}{Abu Awal~Md Shoeb}, \bibinfo{person}{Abubakar Abid}, \bibinfo{person}{Adam Fisch}, \bibinfo{person}{Adam~R. Brown}, \bibinfo{person}{Adam Santoro}, \bibinfo{person}{Aditya Gupta}, \bibinfo{person}{Adrià Garriga-Alonso}, \bibinfo{person}{Agnieszka Kluska}, \bibinfo{person}{Aitor Lewkowycz}, \bibinfo{person}{Akshat Agarwal}, \bibinfo{person}{Alethea Power}, \bibinfo{person}{Alex Ray}, \bibinfo{person}{Alex Warstadt}, \bibinfo{person}{Alexander~W. Kocurek}, \bibinfo{person}{Ali Safaya}, \bibinfo{person}{Ali Tazarv}, \bibinfo{person}{Alice Xiang}, \bibinfo{person}{Alicia Parrish}, \bibinfo{person}{Allen Nie}, \bibinfo{person}{Aman Hussain}, \bibinfo{person}{Amanda Askell}, \bibinfo{person}{Amanda Dsouza}, \bibinfo{person}{Ambrose Slone}, \bibinfo{person}{Ameet Rahane}, \bibinfo{person}{Anantharaman~S. Iyer}, \bibinfo{person}{Anders Andreassen}, \bibinfo{person}{Andrea
  Madotto}, \bibinfo{person}{Andrea Santilli}, \bibinfo{person}{Andreas Stuhlmüller}, \bibinfo{person}{Andrew Dai}, \bibinfo{person}{Andrew La}, \bibinfo{person}{Andrew Lampinen}, \bibinfo{person}{Andy Zou}, \bibinfo{person}{Angela Jiang}, \bibinfo{person}{Angelica Chen}, \bibinfo{person}{Anh Vuong}, \bibinfo{person}{Animesh Gupta}, \bibinfo{person}{Anna Gottardi}, \bibinfo{person}{Antonio Norelli}, \bibinfo{person}{Anu Venkatesh}, \bibinfo{person}{Arash Gholamidavoodi}, \bibinfo{person}{Arfa Tabassum}, \bibinfo{person}{Arul Menezes}, \bibinfo{person}{Arun Kirubarajan}, \bibinfo{person}{Asher Mullokandov}, \bibinfo{person}{Ashish Sabharwal}, \bibinfo{person}{Austin Herrick}, \bibinfo{person}{Avia Efrat}, \bibinfo{person}{Aykut Erdem}, \bibinfo{person}{Ayla Karakaş}, \bibinfo{person}{B.~Ryan Roberts}, \bibinfo{person}{Bao~Sheng Loe}, \bibinfo{person}{Barret Zoph}, \bibinfo{person}{Bartłomiej Bojanowski}, \bibinfo{person}{Batuhan Özyurt}, \bibinfo{person}{Behnam Hedayatnia}, \bibinfo{person}{Behnam
  Neyshabur}, \bibinfo{person}{Benjamin Inden}, \bibinfo{person}{Benno Stein}, \bibinfo{person}{Berk Ekmekci}, \bibinfo{person}{Bill~Yuchen Lin}, \bibinfo{person}{Blake Howald}, \bibinfo{person}{Bryan Orinion}, \bibinfo{person}{Cameron Diao}, \bibinfo{person}{Cameron Dour}, \bibinfo{person}{Catherine Stinson}, \bibinfo{person}{Cedrick Argueta}, \bibinfo{person}{César~Ferri Ramírez}, \bibinfo{person}{Chandan Singh}, \bibinfo{person}{Charles Rathkopf}, \bibinfo{person}{Chenlin Meng}, \bibinfo{person}{Chitta Baral}, \bibinfo{person}{Chiyu Wu}, \bibinfo{person}{Chris Callison-Burch}, \bibinfo{person}{Chris Waites}, \bibinfo{person}{Christian Voigt}, \bibinfo{person}{Christopher~D. Manning}, \bibinfo{person}{Christopher Potts}, \bibinfo{person}{Cindy Ramirez}, \bibinfo{person}{Clara~E. Rivera}, \bibinfo{person}{Clemencia Siro}, \bibinfo{person}{Colin Raffel}, \bibinfo{person}{Courtney Ashcraft}, \bibinfo{person}{Cristina Garbacea}, \bibinfo{person}{Damien Sileo}, \bibinfo{person}{Dan Garrette},
  \bibinfo{person}{Dan Hendrycks}, \bibinfo{person}{Dan Kilman}, \bibinfo{person}{Dan Roth}, \bibinfo{person}{Daniel Freeman}, \bibinfo{person}{Daniel Khashabi}, \bibinfo{person}{Daniel Levy}, \bibinfo{person}{Daniel~Moseguí González}, \bibinfo{person}{Danielle Perszyk}, \bibinfo{person}{Danny Hernandez}, \bibinfo{person}{Danqi Chen}, \bibinfo{person}{Daphne Ippolito}, \bibinfo{person}{Dar Gilboa}, \bibinfo{person}{David Dohan}, \bibinfo{person}{David Drakard}, \bibinfo{person}{David Jurgens}, \bibinfo{person}{Debajyoti Datta}, \bibinfo{person}{Deep Ganguli}, \bibinfo{person}{Denis Emelin}, \bibinfo{person}{Denis Kleyko}, \bibinfo{person}{Deniz Yuret}, \bibinfo{person}{Derek Chen}, \bibinfo{person}{Derek Tam}, \bibinfo{person}{Dieuwke Hupkes}, \bibinfo{person}{Diganta Misra}, \bibinfo{person}{Dilyar Buzan}, \bibinfo{person}{Dimitri~Coelho Mollo}, \bibinfo{person}{Diyi Yang}, \bibinfo{person}{Dong-Ho Lee}, \bibinfo{person}{Dylan Schrader}, \bibinfo{person}{Ekaterina Shutova}, \bibinfo{person}{Ekin~Dogus
  Cubuk}, \bibinfo{person}{Elad Segal}, \bibinfo{person}{Eleanor Hagerman}, \bibinfo{person}{Elizabeth Barnes}, \bibinfo{person}{Elizabeth Donoway}, \bibinfo{person}{Ellie Pavlick}, \bibinfo{person}{Emanuele Rodola}, \bibinfo{person}{Emma Lam}, \bibinfo{person}{Eric Chu}, \bibinfo{person}{Eric Tang}, \bibinfo{person}{Erkut Erdem}, \bibinfo{person}{Ernie Chang}, \bibinfo{person}{Ethan~A. Chi}, \bibinfo{person}{Ethan Dyer}, \bibinfo{person}{Ethan Jerzak}, \bibinfo{person}{Ethan Kim}, \bibinfo{person}{Eunice~Engefu Manyasi}, \bibinfo{person}{Evgenii Zheltonozhskii}, \bibinfo{person}{Fanyue Xia}, \bibinfo{person}{Fatemeh Siar}, \bibinfo{person}{Fernando Martínez-Plumed}, \bibinfo{person}{Francesca Happé}, \bibinfo{person}{Francois Chollet}, \bibinfo{person}{Frieda Rong}, \bibinfo{person}{Gaurav Mishra}, \bibinfo{person}{Genta~Indra Winata}, \bibinfo{person}{Gerard de Melo}, \bibinfo{person}{Germán Kruszewski}, \bibinfo{person}{Giambattista Parascandolo}, \bibinfo{person}{Giorgio Mariani},
  \bibinfo{person}{Gloria Wang}, \bibinfo{person}{Gonzalo Jaimovitch-López}, \bibinfo{person}{Gregor Betz}, \bibinfo{person}{Guy Gur-Ari}, \bibinfo{person}{Hana Galijasevic}, \bibinfo{person}{Hannah Kim}, \bibinfo{person}{Hannah Rashkin}, \bibinfo{person}{Hannaneh Hajishirzi}, \bibinfo{person}{Harsh Mehta}, \bibinfo{person}{Hayden Bogar}, \bibinfo{person}{Henry Shevlin}, \bibinfo{person}{Hinrich Schütze}, \bibinfo{person}{Hiromu Yakura}, \bibinfo{person}{Hongming Zhang}, \bibinfo{person}{Hugh~Mee Wong}, \bibinfo{person}{Ian Ng}, \bibinfo{person}{Isaac Noble}, \bibinfo{person}{Jaap Jumelet}, \bibinfo{person}{Jack Geissinger}, \bibinfo{person}{Jackson Kernion}, \bibinfo{person}{Jacob Hilton}, \bibinfo{person}{Jaehoon Lee}, \bibinfo{person}{Jaime~Fernández Fisac}, \bibinfo{person}{James~B. Simon}, \bibinfo{person}{James Koppel}, \bibinfo{person}{James Zheng}, \bibinfo{person}{James Zou}, \bibinfo{person}{Jan Kocoń}, \bibinfo{person}{Jana Thompson}, \bibinfo{person}{Janelle Wingfield}, \bibinfo{person}{Jared
  Kaplan}, \bibinfo{person}{Jarema Radom}, \bibinfo{person}{Jascha Sohl-Dickstein}, \bibinfo{person}{Jason Phang}, \bibinfo{person}{Jason Wei}, \bibinfo{person}{Jason Yosinski}, \bibinfo{person}{Jekaterina Novikova}, \bibinfo{person}{Jelle Bosscher}, \bibinfo{person}{Jennifer Marsh}, \bibinfo{person}{Jeremy Kim}, \bibinfo{person}{Jeroen Taal}, \bibinfo{person}{Jesse Engel}, \bibinfo{person}{Jesujoba Alabi}, \bibinfo{person}{Jiacheng Xu}, \bibinfo{person}{Jiaming Song}, \bibinfo{person}{Jillian Tang}, \bibinfo{person}{Joan Waweru}, \bibinfo{person}{John Burden}, \bibinfo{person}{John Miller}, \bibinfo{person}{John~U. Balis}, \bibinfo{person}{Jonathan Batchelder}, \bibinfo{person}{Jonathan Berant}, \bibinfo{person}{Jörg Frohberg}, \bibinfo{person}{Jos Rozen}, \bibinfo{person}{Jose Hernandez-Orallo}, \bibinfo{person}{Joseph Boudeman}, \bibinfo{person}{Joseph Guerr}, \bibinfo{person}{Joseph Jones}, \bibinfo{person}{Joshua~B. Tenenbaum}, \bibinfo{person}{Joshua~S. Rule}, \bibinfo{person}{Joyce Chua},
  \bibinfo{person}{Kamil Kanclerz}, \bibinfo{person}{Karen Livescu}, \bibinfo{person}{Karl Krauth}, \bibinfo{person}{Karthik Gopalakrishnan}, \bibinfo{person}{Katerina Ignatyeva}, \bibinfo{person}{Katja Markert}, \bibinfo{person}{Kaustubh~D. Dhole}, \bibinfo{person}{Kevin Gimpel}, \bibinfo{person}{Kevin Omondi}, \bibinfo{person}{Kory Mathewson}, \bibinfo{person}{Kristen Chiafullo}, \bibinfo{person}{Ksenia Shkaruta}, \bibinfo{person}{Kumar Shridhar}, \bibinfo{person}{Kyle McDonell}, \bibinfo{person}{Kyle Richardson}, \bibinfo{person}{Laria Reynolds}, \bibinfo{person}{Leo Gao}, \bibinfo{person}{Li Zhang}, \bibinfo{person}{Liam Dugan}, \bibinfo{person}{Lianhui Qin}, \bibinfo{person}{Lidia Contreras-Ochando}, \bibinfo{person}{Louis-Philippe Morency}, \bibinfo{person}{Luca Moschella}, \bibinfo{person}{Lucas Lam}, \bibinfo{person}{Lucy Noble}, \bibinfo{person}{Ludwig Schmidt}, \bibinfo{person}{Luheng He}, \bibinfo{person}{Luis~Oliveros Colón}, \bibinfo{person}{Luke Metz}, \bibinfo{person}{Lütfi~Kerem Şenel},
  \bibinfo{person}{Maarten Bosma}, \bibinfo{person}{Maarten Sap}, \bibinfo{person}{Maartje ter Hoeve}, \bibinfo{person}{Maheen Farooqi}, \bibinfo{person}{Manaal Faruqui}, \bibinfo{person}{Mantas Mazeika}, \bibinfo{person}{Marco Baturan}, \bibinfo{person}{Marco Marelli}, \bibinfo{person}{Marco Maru}, \bibinfo{person}{Maria Jose~Ramírez Quintana}, \bibinfo{person}{Marie Tolkiehn}, \bibinfo{person}{Mario Giulianelli}, \bibinfo{person}{Martha Lewis}, \bibinfo{person}{Martin Potthast}, \bibinfo{person}{Matthew~L. Leavitt}, \bibinfo{person}{Matthias Hagen}, \bibinfo{person}{Mátyás Schubert}, \bibinfo{person}{Medina~Orduna Baitemirova}, \bibinfo{person}{Melody Arnaud}, \bibinfo{person}{Melvin McElrath}, \bibinfo{person}{Michael~A. Yee}, \bibinfo{person}{Michael Cohen}, \bibinfo{person}{Michael Gu}, \bibinfo{person}{Michael Ivanitskiy}, \bibinfo{person}{Michael Starritt}, \bibinfo{person}{Michael Strube}, \bibinfo{person}{Michał Swędrowski}, \bibinfo{person}{Michele Bevilacqua}, \bibinfo{person}{Michihiro
  Yasunaga}, \bibinfo{person}{Mihir Kale}, \bibinfo{person}{Mike Cain}, \bibinfo{person}{Mimee Xu}, \bibinfo{person}{Mirac Suzgun}, \bibinfo{person}{Mitch Walker}, \bibinfo{person}{Mo Tiwari}, \bibinfo{person}{Mohit Bansal}, \bibinfo{person}{Moin Aminnaseri}, \bibinfo{person}{Mor Geva}, \bibinfo{person}{Mozhdeh Gheini}, \bibinfo{person}{Mukund~Varma T}, \bibinfo{person}{Nanyun Peng}, \bibinfo{person}{Nathan~A. Chi}, \bibinfo{person}{Nayeon Lee}, \bibinfo{person}{Neta Gur-Ari Krakover}, \bibinfo{person}{Nicholas Cameron}, \bibinfo{person}{Nicholas Roberts}, \bibinfo{person}{Nick Doiron}, \bibinfo{person}{Nicole Martinez}, \bibinfo{person}{Nikita Nangia}, \bibinfo{person}{Niklas Deckers}, \bibinfo{person}{Niklas Muennighoff}, \bibinfo{person}{Nitish~Shirish Keskar}, \bibinfo{person}{Niveditha~S. Iyer}, \bibinfo{person}{Noah Constant}, \bibinfo{person}{Noah Fiedel}, \bibinfo{person}{Nuan Wen}, \bibinfo{person}{Oliver Zhang}, \bibinfo{person}{Omar Agha}, \bibinfo{person}{Omar Elbaghdadi}, \bibinfo{person}{Omer
  Levy}, \bibinfo{person}{Owain Evans}, \bibinfo{person}{Pablo Antonio~Moreno Casares}, \bibinfo{person}{Parth Doshi}, \bibinfo{person}{Pascale Fung}, \bibinfo{person}{Paul~Pu Liang}, \bibinfo{person}{Paul Vicol}, \bibinfo{person}{Pegah Alipoormolabashi}, \bibinfo{person}{Peiyuan Liao}, \bibinfo{person}{Percy Liang}, \bibinfo{person}{Peter Chang}, \bibinfo{person}{Peter Eckersley}, \bibinfo{person}{Phu~Mon Htut}, \bibinfo{person}{Pinyu Hwang}, \bibinfo{person}{Piotr Miłkowski}, \bibinfo{person}{Piyush Patil}, \bibinfo{person}{Pouya Pezeshkpour}, \bibinfo{person}{Priti Oli}, \bibinfo{person}{Qiaozhu Mei}, \bibinfo{person}{Qing Lyu}, \bibinfo{person}{Qinlang Chen}, \bibinfo{person}{Rabin Banjade}, \bibinfo{person}{Rachel~Etta Rudolph}, \bibinfo{person}{Raefer Gabriel}, \bibinfo{person}{Rahel Habacker}, \bibinfo{person}{Ramon Risco}, \bibinfo{person}{Raphaël Millière}, \bibinfo{person}{Rhythm Garg}, \bibinfo{person}{Richard Barnes}, \bibinfo{person}{Rif~A. Saurous}, \bibinfo{person}{Riku Arakawa},
  \bibinfo{person}{Robbe Raymaekers}, \bibinfo{person}{Robert Frank}, \bibinfo{person}{Rohan Sikand}, \bibinfo{person}{Roman Novak}, \bibinfo{person}{Roman Sitelew}, \bibinfo{person}{Ronan LeBras}, \bibinfo{person}{Rosanne Liu}, \bibinfo{person}{Rowan Jacobs}, \bibinfo{person}{Rui Zhang}, \bibinfo{person}{Ruslan Salakhutdinov}, \bibinfo{person}{Ryan Chi}, \bibinfo{person}{Ryan Lee}, \bibinfo{person}{Ryan Stovall}, \bibinfo{person}{Ryan Teehan}, \bibinfo{person}{Rylan Yang}, \bibinfo{person}{Sahib Singh}, \bibinfo{person}{Saif~M. Mohammad}, \bibinfo{person}{Sajant Anand}, \bibinfo{person}{Sam Dillavou}, \bibinfo{person}{Sam Shleifer}, \bibinfo{person}{Sam Wiseman}, \bibinfo{person}{Samuel Gruetter}, \bibinfo{person}{Samuel~R. Bowman}, \bibinfo{person}{Samuel~S. Schoenholz}, \bibinfo{person}{Sanghyun Han}, \bibinfo{person}{Sanjeev Kwatra}, \bibinfo{person}{Sarah~A. Rous}, \bibinfo{person}{Sarik Ghazarian}, \bibinfo{person}{Sayan Ghosh}, \bibinfo{person}{Sean Casey}, \bibinfo{person}{Sebastian Bischoff},
  \bibinfo{person}{Sebastian Gehrmann}, \bibinfo{person}{Sebastian Schuster}, \bibinfo{person}{Sepideh Sadeghi}, \bibinfo{person}{Shadi Hamdan}, \bibinfo{person}{Sharon Zhou}, \bibinfo{person}{Shashank Srivastava}, \bibinfo{person}{Sherry Shi}, \bibinfo{person}{Shikhar Singh}, \bibinfo{person}{Shima Asaadi}, \bibinfo{person}{Shixiang~Shane Gu}, \bibinfo{person}{Shubh Pachchigar}, \bibinfo{person}{Shubham Toshniwal}, \bibinfo{person}{Shyam Upadhyay}, \bibinfo{person}{Shyamolima}, \bibinfo{person}{Debnath}, \bibinfo{person}{Siamak Shakeri}, \bibinfo{person}{Simon Thormeyer}, \bibinfo{person}{Simone Melzi}, \bibinfo{person}{Siva Reddy}, \bibinfo{person}{Sneha~Priscilla Makini}, \bibinfo{person}{Soo-Hwan Lee}, \bibinfo{person}{Spencer Torene}, \bibinfo{person}{Sriharsha Hatwar}, \bibinfo{person}{Stanislas Dehaene}, \bibinfo{person}{Stefan Divic}, \bibinfo{person}{Stefano Ermon}, \bibinfo{person}{Stella Biderman}, \bibinfo{person}{Stephanie Lin}, \bibinfo{person}{Stephen Prasad}, \bibinfo{person}{Steven~T.
  Piantadosi}, \bibinfo{person}{Stuart~M. Shieber}, \bibinfo{person}{Summer Misherghi}, \bibinfo{person}{Svetlana Kiritchenko}, \bibinfo{person}{Swaroop Mishra}, \bibinfo{person}{Tal Linzen}, \bibinfo{person}{Tal Schuster}, \bibinfo{person}{Tao Li}, \bibinfo{person}{Tao Yu}, \bibinfo{person}{Tariq Ali}, \bibinfo{person}{Tatsu Hashimoto}, \bibinfo{person}{Te-Lin Wu}, \bibinfo{person}{Théo Desbordes}, \bibinfo{person}{Theodore Rothschild}, \bibinfo{person}{Thomas Phan}, \bibinfo{person}{Tianle Wang}, \bibinfo{person}{Tiberius Nkinyili}, \bibinfo{person}{Timo Schick}, \bibinfo{person}{Timofei Kornev}, \bibinfo{person}{Titus Tunduny}, \bibinfo{person}{Tobias Gerstenberg}, \bibinfo{person}{Trenton Chang}, \bibinfo{person}{Trishala Neeraj}, \bibinfo{person}{Tushar Khot}, \bibinfo{person}{Tyler Shultz}, \bibinfo{person}{Uri Shaham}, \bibinfo{person}{Vedant Misra}, \bibinfo{person}{Vera Demberg}, \bibinfo{person}{Victoria Nyamai}, \bibinfo{person}{Vikas Raunak}, \bibinfo{person}{Vinay Ramasesh},
  \bibinfo{person}{Vinay~Uday Prabhu}, \bibinfo{person}{Vishakh Padmakumar}, \bibinfo{person}{Vivek Srikumar}, \bibinfo{person}{William Fedus}, \bibinfo{person}{William Saunders}, \bibinfo{person}{William Zhang}, \bibinfo{person}{Wout Vossen}, \bibinfo{person}{Xiang Ren}, \bibinfo{person}{Xiaoyu Tong}, \bibinfo{person}{Xinran Zhao}, \bibinfo{person}{Xinyi Wu}, \bibinfo{person}{Xudong Shen}, \bibinfo{person}{Yadollah Yaghoobzadeh}, \bibinfo{person}{Yair Lakretz}, \bibinfo{person}{Yangqiu Song}, \bibinfo{person}{Yasaman Bahri}, \bibinfo{person}{Yejin Choi}, \bibinfo{person}{Yichi Yang}, \bibinfo{person}{Yiding Hao}, \bibinfo{person}{Yifu Chen}, \bibinfo{person}{Yonatan Belinkov}, \bibinfo{person}{Yu Hou}, \bibinfo{person}{Yufang Hou}, \bibinfo{person}{Yuntao Bai}, \bibinfo{person}{Zachary Seid}, \bibinfo{person}{Zhuoye Zhao}, \bibinfo{person}{Zijian Wang}, \bibinfo{person}{Zijie~J. Wang}, \bibinfo{person}{Zirui Wang}, {and} \bibinfo{person}{Ziyi Wu}.} \bibinfo{year}{2023}\natexlab{}.
\newblock \bibinfo{title}{Beyond the Imitation Game: Quantifying and extrapolating the capabilities of language models}.
\newblock
\newblock
\showeprint[arxiv]{2206.04615}~[cs.CL]


\bibitem[Thoppilan et~al\mbox{.}(2022)]%
        {thoppilan2022lamda}
\bibfield{author}{\bibinfo{person}{Romal Thoppilan}, \bibinfo{person}{Daniel~De Freitas}, \bibinfo{person}{Jamie Hall}, \bibinfo{person}{Noam Shazeer}, \bibinfo{person}{Apoorv Kulshreshtha}, \bibinfo{person}{Heng-Tze Cheng}, \bibinfo{person}{Alicia Jin}, \bibinfo{person}{Taylor Bos}, \bibinfo{person}{Leslie Baker}, \bibinfo{person}{Yu Du}, \bibinfo{person}{YaGuang Li}, \bibinfo{person}{Hongrae Lee}, \bibinfo{person}{Huaixiu~Steven Zheng}, \bibinfo{person}{Amin Ghafouri}, \bibinfo{person}{Marcelo Menegali}, \bibinfo{person}{Yanping Huang}, \bibinfo{person}{Maxim Krikun}, \bibinfo{person}{Dmitry Lepikhin}, \bibinfo{person}{James Qin}, \bibinfo{person}{Dehao Chen}, \bibinfo{person}{Yuanzhong Xu}, \bibinfo{person}{Zhifeng Chen}, \bibinfo{person}{Adam Roberts}, \bibinfo{person}{Maarten Bosma}, \bibinfo{person}{Vincent Zhao}, \bibinfo{person}{Yanqi Zhou}, \bibinfo{person}{Chung-Ching Chang}, \bibinfo{person}{Igor Krivokon}, \bibinfo{person}{Will Rusch}, \bibinfo{person}{Marc Pickett}, \bibinfo{person}{Pranesh
  Srinivasan}, \bibinfo{person}{Laichee Man}, \bibinfo{person}{Kathleen Meier-Hellstern}, \bibinfo{person}{Meredith~Ringel Morris}, \bibinfo{person}{Tulsee Doshi}, \bibinfo{person}{Renelito~Delos Santos}, \bibinfo{person}{Toju Duke}, \bibinfo{person}{Johnny Soraker}, \bibinfo{person}{Ben Zevenbergen}, \bibinfo{person}{Vinodkumar Prabhakaran}, \bibinfo{person}{Mark Diaz}, \bibinfo{person}{Ben Hutchinson}, \bibinfo{person}{Kristen Olson}, \bibinfo{person}{Alejandra Molina}, \bibinfo{person}{Erin Hoffman-John}, \bibinfo{person}{Josh Lee}, \bibinfo{person}{Lora Aroyo}, \bibinfo{person}{Ravi Rajakumar}, \bibinfo{person}{Alena Butryna}, \bibinfo{person}{Matthew Lamm}, \bibinfo{person}{Viktoriya Kuzmina}, \bibinfo{person}{Joe Fenton}, \bibinfo{person}{Aaron Cohen}, \bibinfo{person}{Rachel Bernstein}, \bibinfo{person}{Ray Kurzweil}, \bibinfo{person}{Blaise Aguera-Arcas}, \bibinfo{person}{Claire Cui}, \bibinfo{person}{Marian Croak}, \bibinfo{person}{Ed Chi}, {and} \bibinfo{person}{Quoc Le}.}
  \bibinfo{year}{2022}\natexlab{}.
\newblock \bibinfo{title}{LaMDA: Language Models for Dialog Applications}.
\newblock
\newblock
\showeprint[arxiv]{2201.08239}~[cs.CL]


\bibitem[Vakulenko et~al\mbox{.}(2021)]%
        {DBLP:journals/tois/VakulenkoKR21}
\bibfield{author}{\bibinfo{person}{Svitlana Vakulenko}, \bibinfo{person}{Evangelos Kanoulas}, {and} \bibinfo{person}{Maarten de Rijke}.} \bibinfo{year}{2021}\natexlab{}.
\newblock \showarticletitle{A Large-scale Analysis of Mixed Initiative in Information-Seeking Dialogues for Conversational Search}.
\newblock \bibinfo{journal}{\emph{{ACM} Trans. Inf. Syst.}} \bibinfo{volume}{39}, \bibinfo{number}{4} (\bibinfo{year}{2021}), \bibinfo{pages}{49:1--49:32}.
\newblock


\bibitem[Wang et~al\mbox{.}(2020)]%
        {wang2020superglue}
\bibfield{author}{\bibinfo{person}{Alex Wang}, \bibinfo{person}{Yada Pruksachatkun}, \bibinfo{person}{Nikita Nangia}, \bibinfo{person}{Amanpreet Singh}, \bibinfo{person}{Julian Michael}, \bibinfo{person}{Felix Hill}, \bibinfo{person}{Omer Levy}, {and} \bibinfo{person}{Samuel~R. Bowman}.} \bibinfo{year}{2020}\natexlab{}.
\newblock \bibinfo{title}{SuperGLUE: A Stickier Benchmark for General-Purpose Language Understanding Systems}.
\newblock
\newblock
\showeprint[arxiv]{1905.00537}~[cs.CL]


\bibitem[Yang and Soboroff(2016)]%
        {yang2016trec}
\bibfield{author}{\bibinfo{person}{Grace~Hui Yang} {and} \bibinfo{person}{Ian Soboroff}.} \bibinfo{year}{2016}\natexlab{}.
\newblock \showarticletitle{TREC 2016 Dynamic Domain Track Overview.}. In \bibinfo{booktitle}{\emph{Text REtrieval Conference ({TREC})}}. \bibinfo{publisher}{NIST}.
\newblock


\bibitem[Yang et~al\mbox{.}(2017)]%
        {DBLP:conf/trec/YangTS17}
\bibfield{author}{\bibinfo{person}{Grace~Hui Yang}, \bibinfo{person}{Zhiwen Tang}, {and} \bibinfo{person}{Ian Soboroff}.} \bibinfo{year}{2017}\natexlab{}.
\newblock \showarticletitle{{TREC} 2017 Dynamic Domain Track Overview}. In \bibinfo{booktitle}{\emph{Text REtrieval Conference ({TREC})}}. \bibinfo{publisher}{NIST}.
\newblock


\bibitem[Yang et~al\mbox{.}(2015)]%
        {DBLP:conf/trec/0001FS15}
\bibfield{author}{\bibinfo{person}{Hui Yang}, \bibinfo{person}{John~R. Frank}, {and} \bibinfo{person}{Ian Soboroff}.} \bibinfo{year}{2015}\natexlab{}.
\newblock \showarticletitle{{TREC} 2015 Dynamic Domain Track Overview}. In \bibinfo{booktitle}{\emph{Text REtrieval Conference ({TREC})}}. \bibinfo{publisher}{NIST}.
\newblock


\bibitem[Zellers et~al\mbox{.}(2019)]%
        {zellers2019hellaswag}
\bibfield{author}{\bibinfo{person}{Rowan Zellers}, \bibinfo{person}{Ari Holtzman}, \bibinfo{person}{Yonatan Bisk}, \bibinfo{person}{Ali Farhadi}, {and} \bibinfo{person}{Yejin Choi}.} \bibinfo{year}{2019}\natexlab{}.
\newblock \bibinfo{title}{HellaSwag: Can a Machine Really Finish Your Sentence?}
\newblock
\newblock
\showeprint[arxiv]{1905.07830}~[cs.CL]


\bibitem[Zheng et~al\mbox{.}(2023)]%
        {zheng2023judging}
\bibfield{author}{\bibinfo{person}{Lianmin Zheng}, \bibinfo{person}{Wei-Lin Chiang}, \bibinfo{person}{Ying Sheng}, \bibinfo{person}{Siyuan Zhuang}, \bibinfo{person}{Zhanghao Wu}, \bibinfo{person}{Yonghao Zhuang}, \bibinfo{person}{Zi Lin}, \bibinfo{person}{Zhuohan Li}, \bibinfo{person}{Dacheng Li}, \bibinfo{person}{Eric.~P Xing}, \bibinfo{person}{Hao Zhang}, \bibinfo{person}{Joseph~E. Gonzalez}, {and} \bibinfo{person}{Ion Stoica}.} \bibinfo{year}{2023}\natexlab{}.
\newblock \bibinfo{title}{Judging LLM-as-a-judge with MT-Bench and Chatbot Arena}.
\newblock
\newblock
\showeprint[arxiv]{2306.05685}~[cs.CL]


\end{thebibliography}
\balance
\clearpage
\appendix
\include{appendix-example.tex}

\end{document}